\documentclass[fleqn,usenatbib]{mnras}

\usepackage{newtxtext,newtxmath}

\usepackage[T1]{fontenc}
\usepackage{ae,aecompl}

\usepackage{graphicx}	
\usepackage{amsmath}	
\usepackage{amssymb}	
\usepackage{multicol}
\usepackage{bm}
\usepackage{booktabs}

\newcommand{\CIV}{C\textsc{iv}$\lambda \lambda 1548,1550$ }
\newcommand{\HeII}{He\textsc{ii}$\lambda1640$ }
\newcommand{\secpoint}{\mbox{$''\mskip-7.6mu.\,$}}
\newcommand{\angstrom}{\mbox{\normalfont\AA}}

\title[Massive Stars and Ionized Gas at High Redshift]{The MOSDEF-LRIS Survey: The Interplay Between Massive Stars and Ionized Gas in High-Redshift Star-Forming Galaxies$^{1}$}

\author[M. W. Topping et al.]{Michael W. Topping,$^{2}$\thanks{E-mail: mtopping@astro.ucla.edu}
Alice E. Shapley,$^{2}$
Naveen A. Reddy,$^{3}$
Ryan L. Sanders,$^{4}$\newauthor
Alison L. Coil,$^{5}$
Mariska Kriek,$^{6}$
Bahram Mobasher,$^{3}$	
Brian Siana,$^{3}$
\\
$^{1}$Based on data obtained at the W.M. Keck Observatory, which is operated as a scientific partnership among the California Institute of Technology, \\ the University of California,  and the National Aeronautics and Space Administration, and was made possible by the generous financial support \\ of the W.M. Keck Foundation.\\
$^{2}$Department of Physics and Astronomy, University of California, Los Angeles, 430 Portola Plaza, Los Angeles, CA 90095, USA\\
$^{3}$Department of Physics and Astronomy, University of California, Riverside, 900 University Avenue, Riverside, CA 92521, USA\\
$^{4}$Department of Physics, University of California, Davis, 1 Shields Avenue, Davis, CA 95616, USA\\
$^{5}$Center for Astrophysics and Space Sciences, Department of Physics, University of California, San Diego, 9500 Gilman Drive., La Jolla, CA 92093, USA\\
$^{6}$Astronomy Department, University of California at Berkeley, Berkeley, CA 94720, USA
}

\begin{document}
\label{firstpage}
\pagerange{\pageref{firstpage}--\pageref{lastpage}}
\maketitle

\begin{abstract}

We present a joint analysis of rest-UV and rest-optical spectra obtained using Keck/LRIS and Keck/MOSFIRE for a sample of 62 star-forming galaxies at $z\sim2.3$. We divide our sample into two bins based on their location in the [OIII]5007/H$\beta$ vs. [NII]6584/H$\alpha$ BPT diagram, and perform the first differential study of the rest-UV properties of massive ionizing stars as a function of rest-optical emission-line ratios. Fitting BPASS stellar population synthesis models, including nebular continuum emission, to our rest-UV composite spectra, we find that high-redshift galaxies offset towards higher [OIII]$\lambda 5007$/H$\beta$ and [NII]$\lambda 6584$/H$\alpha$ have younger ages ($\log(\textrm{ Age/yr})=7.20^{+0.57}_{-0.20}$) and lower stellar metallicities ($Z_*=0.0010^{+0.0011}_{-0.0003}$) resulting in a harder ionizing spectrum, compared to the galaxies in our sample that lie on the local BPT star-forming sequence ($\log(\textrm{Age/yr})=8.57^{+0.88}_{-0.84}$, $Z_*=0.0019^{+0.0006}_{-0.0006}$).  Additionally, we find that the offset galaxies have an ionization parameter of $\log(U)=-3.04^{+0.06}_{-0.11}$ and nebular metallicity of ($12+\log(\textrm{ O/H})=8.40^{+0.06}_{-0.07}$), and the non-offset galaxies have an ionization parameter of $\log(U)=-3.11^{+0.08}_{-0.08}$ and nebular metallicity of $12+\log(\textrm{ O/H})=8.30^{+0.05}_{-0.06}$. The stellar and nebular metallicities derived for our sample imply that the galaxies offset from the local BPT relation are more $\alpha$-enhanced ($7.28^{+2.52}_{-2.82}\textrm{ O/Fe}_{\odot}$) compared to those consistent with the local sequence ($3.04^{+0.95}_{-0.54}\textrm{ O/Fe}_{\odot}$). However, even galaxies that are entirely consistent with the local nebular excitation sequence appear to be $\alpha$-enhanced -- in contrast with typical local systems.  Such differences must be considered when estimating gas-phase oxygen abundances at high redshift based on strong emission-line ratios. Specifically, a similarity in the location of high-redshift and local galaxies in the BPT diagram may not be indicative of a similarity in their physical properties.

\end{abstract}

\begin{keywords}
galaxies: evolution -- galaxies: ISM -- galaxies: high-redshift
\end{keywords}

\section{Introduction} 
\label{sec:intro}


Rest-optical spectroscopy is a powerful tool that can be used to determine a wealth of information on the physical conditions within the interstellar medium (ISM) of galaxies.  Measurements of optical nebular emission lines from local star-forming galaxies demonstrate that they trace a tight sequence of increasing [NII]$\lambda 6584$/H$\alpha$ and decreasing [OIII]$\lambda 5007$/H$\beta$ emission-line ratios \citep[e.g.,][]{Veilleux1987,Kauffmann2003}.  The observed variation in emission-line ratios along the star-forming sequence reflects the increasing oxygen abundance and stellar mass and decreasing H\textsc{ii}-region excitation of its constituent galaxies \citep{Masters2016}. Early observations with Keck/NIRSPEC suggested possible differences in the emission-line properties of high-redshift galaxies in the [OIII]$\lambda 5007$/H$\beta$ vs. [NII]$\lambda 6584$/H$\alpha$ ``BPT'' diagram \citep{Baldwin1981, Shapley2005, Erb2006, Liu2008}. New, statistical samples from the MOSFIRE Deep Evolution Field \citep[MOSDEF;][]{Kriek2015} survey and the Keck Baryonic Structure Survey \citep[KBSS;][]{Steidel2014} show that typical high-redshift galaxies are offset towards higher [OIII]$\lambda 5007$/H$\beta$ and/or [NII]$\lambda 6584$/H$\alpha$ on average relative to local galaxies.  

\begin{figure*}
    \includegraphics[width=0.8\linewidth]{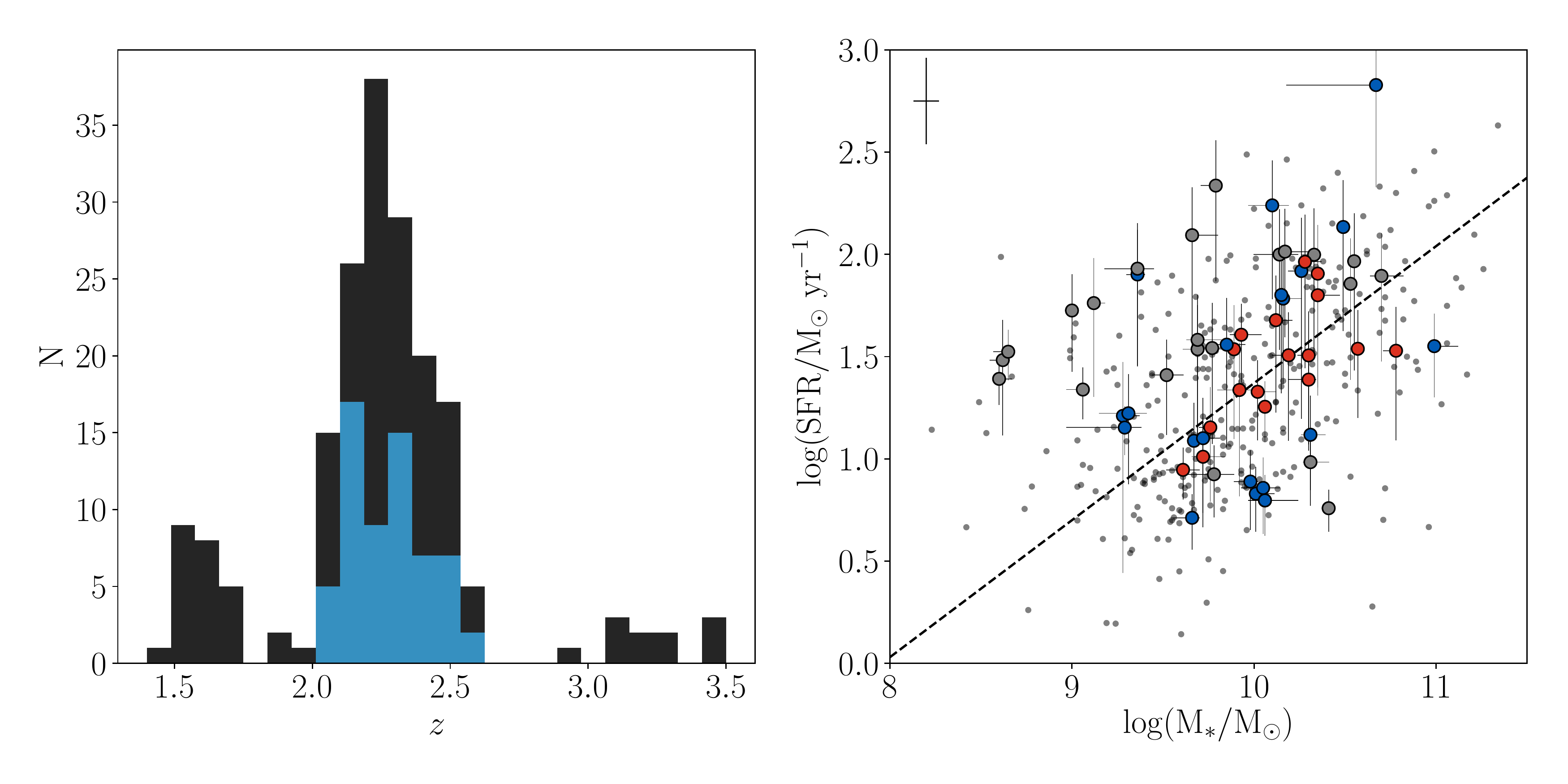}
    \caption{Left: Redshift histogram for all objects with redshifts measured from LRIS spectra (black), totalling 188 galaxies. The blue redshift histogram comprises all objects within $2.09\le z \le 2.61$  ($z_{\textrm{ med}}=2.28$) with all four BPT lines (H$\beta$, [OIII]$\lambda 5007$, H$\alpha$, [NII]$\lambda6584$) detected at $\ge 3\sigma$ in MOSFIRE spectra from the MOSDEF survey, totalling 62 galaxies.  Right: SFR calculated from the dust-corrected Balmer lines vs. $\textrm{ M}_*$ for all objects with LRIS redshifts at $2.0\le z \le 2.7$ (large circles). Blue and red points indicate galaxies included, respectively, in the \textit{high} and \textit{low} composite spectra described in Section~\ref{sec:results}.  Galaxies from the MOSDEF survey within $2.0\le z \le 2.7$ that have both H$\alpha$ and H$\beta$ detected with $\ge3\sigma$ are depicted by small grey points. The median errorbar is shown in the top left corner. The dashed line shows the SFR-$\textrm{ M}_*$ relation of $z\sim2.3$ galaxies from the MOSDEF survey calculated by \citet{Sanders2018}.}
    \label{fig:bptsample}
\end{figure*}

There are many possible causes for this observed difference between local and $z\sim2$ galaxies, including higher ionization parameters, harder ionizing spectra at fixed nebular metallicity, higher densities, variations in gas-phase abundance patterns, and enhanced contributions from AGNs and shocks at high redshift \citep[see e.g.,][ for a review]{Kewley2013}.  Early results from the MOSDEF survey suggested that the offset of high-redshift galaxies on the BPT diagram is caused in part by the order-of-magnitude higher physical density in $z\sim2$ star-forming regions, but is primarily a result of an enhanced N/O ratio abundance at fixed oxygen abundance in offset $z\sim2$ star-forming galaxies relative to local systems \citep{Masters2014,Shapley2015,Sanders2016a}. Furthermore, there is evidence that the BPT offset is strongest among low-mass, young galaxies \citep{Shapley2015,Strom2017}.  Results from KBSS were used to argue instead that the observed offset is more likely driven by a harder stellar ionizing spectrum at fixed nebular metallicity, which can also explain at least some of the observed emission-line patterns \citep{Steidel2016, Strom2017}.Recently, updated results from the MOSDEF survey corroborate these results favoring a harder stellar ionizing spectrum at fixed nebular metallicity \citep{Sanders2019, Shapley2019}, which arises naturally due to the super-solar O/Fe values of the massive ionizing stars that excite the H\textsc{ii} regions in these $z\sim 2$ star-forming galaxies. Such $\alpha$-enhancement would naturally exist in high-redshift galaxies due to rapid formation timescales, resulting in enrichment by a larger fraction of Type II relative to Type Ia supernova explosions.

In star-forming galaxies, massive stars are the predominant sources of ionizing radiation driving the nebular emission lines included in the BPT diagram.  As such, studying the properties of massive stars enables us to address the origin of the observed rest-optical spectroscopic differences between local and high-redshift galaxies. The formation and evolution of massive stars is intimately linked with the evolving properties of the ionized ISM. Specifically, the formation of massive stars is driven by the accretion of gas onto galaxies, and, in turn, massive stars regulate the chemical enrichment of the ISM by driving galaxy-scale outflows, and polluting the ISM when they explode as core-collapse supernovae. Additionally, due to the short-lived nature of these stars, they provide a probe of star-forming galaxies on timescales shorter to or equal to the typical dynamical timescale. One avenue for studying the properties of the massive star populations in high-redshift galaxies is directly observing their light using rest-UV spectroscopy.

\begin{figure*}
    \centering
    \includegraphics[width=0.9\linewidth]{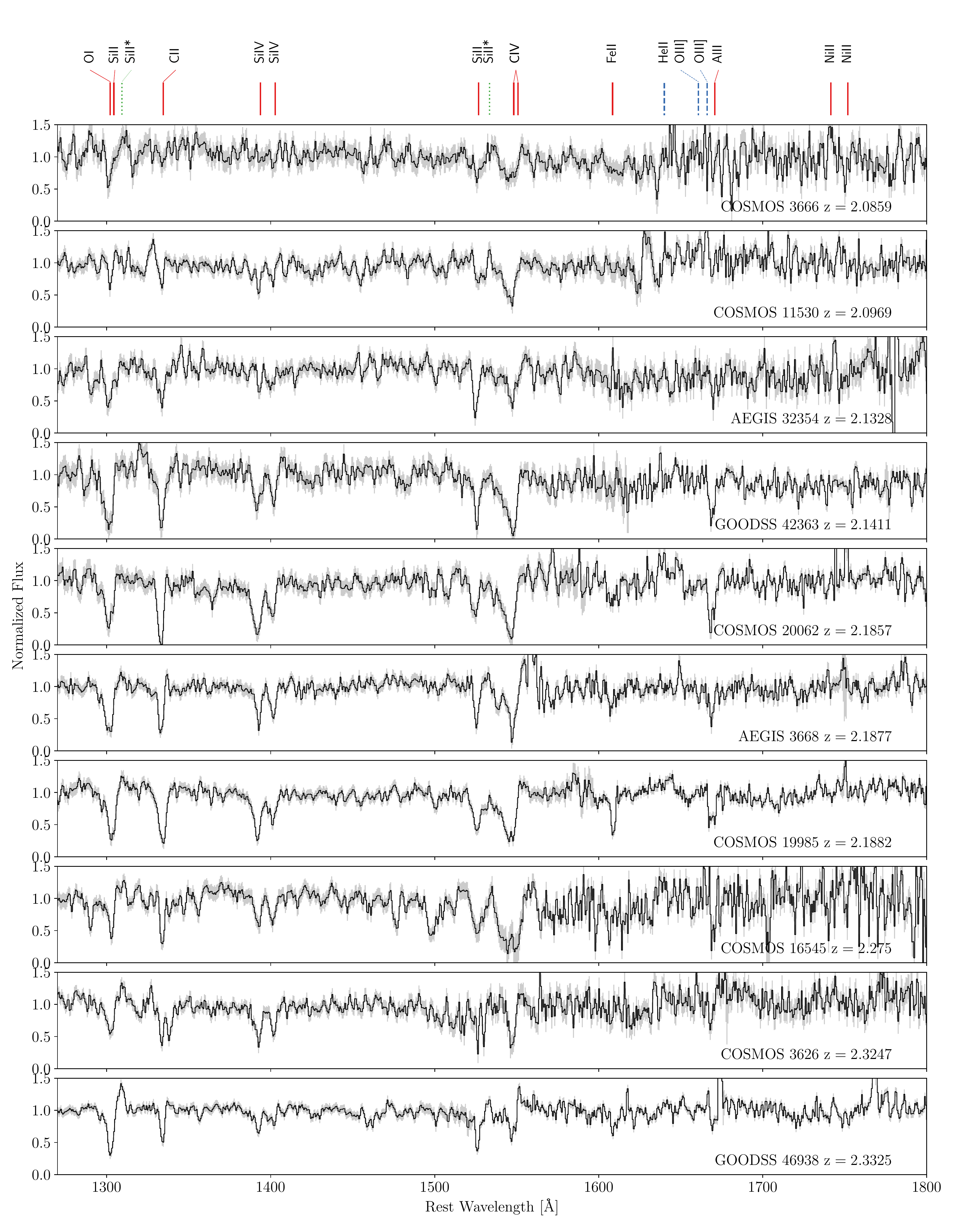}
    \caption{Ten continuum-normalized individual rest-UV spectra from our sample ordered by redshift.  These spectra have the highest continuum signal-to-noise ratio (SNR) from our sample with median $\textrm{ SNR/pixel} = 7$ ($4.5 \le \textrm{ SNR/pixel} \le 12$), measured over the wavelength range $1425\angstrom \le \lambda \le 1500 \textrm{ \angstrom}$.  Labels on the top of the figure indicate several spectral features including stellar absorption lines (solid red lines), nebular emission lines (dashed dark blue lines), and fine structure lines (dotted green lines). At these redshifts, the dichroic cutoff between the red- and blue-side spectra occurs at a typical rest-frame wavelength of $\sim 1500\textrm{ \angstrom}$.  The $1\sigma$ error spectrum is depicted by the shaded region surrounding each spectrum.}
    \label{fig:bptindividual}
\end{figure*}

Rest-UV spectra of star-forming galaxies contain many features tracing the massive, young stars that supply the ionizing luminosity exciting the gas in star-forming regions. These features, such as the \CIV \citep{Crowther2006, Leitherer2001} and \HeII \citep{Brinchmann2008} stellar wind lines, and a host of stellar photospheric features \citep{Rix2004}, provide information on the population of massive stars. In particular, using a given set of model assumptions, these features can be used to establish the form of the initial mass function (IMF), the abundance of Wolf-Rayet (WR) stars, and the nature of the ionizing spectrum in star-forming regions.  The features of rest-UV spectra have also been used to estimate stellar abundances (i.e., Fe/H) in high-redshift galaxies. \citet{Halliday2008} used the Fe\textsc{iii}-sensitive $1978\textrm{ \angstrom}$ index defined by \citet{Rix2004} to measure a stellar metallicity of $Z_*/Z_{\odot}=0.267$ in a composite spectrum of 75 $z\sim2$ star-forming galaxies. \citet{Sommariva2012} employed a similar approach, and investigated new photospheric absorption line indicators suitable as calibrations of the stellar metallicity in high-redshift galaxies.  They applied these calibrations to the rest-UV spectra of nine $z\sim3.3$ individual galaxies, and one composite spectrum to construct the $z\ge2.5$ $M_*$-$Z_*$ relation.  Compared to the previously mentioned works, \citet{Cullen2019} instead fit models to the full rest-UV spectrum, an approach that uses all of the stellar-metallicity sensitive spectral features simultaneously. They applied this method to composite spectra to constrain the stellar metallicity of star-forming galaxies spanning a redshift range of $2.5< z < 5.0$ and a stellar mass range of $8.5 < \log(M_*/M_{\odot}) < 10.2$.

Expanding on previous work, recent studies have made use of rest-UV spectra in combination with rest-optical spectra of high-redshift galaxies \citep{Steidel2016, Chisholm2019}.  Using composite rest-UV and rest-optical spectra of 30 star-forming galaxies at $z\sim2.4$, \citet{Steidel2016} found that the observed properties constrained by their composite spectra can be reproduced only by models that include binary stars, have low stellar metallicities ($Z_*/Z_{\odot}\sim0.1$) and moderate nebular metallicities ($Z_{\textrm{ neb}}/Z_{\odot}\sim0.5$). These results indicate $\alpha$-enhancement for the $z\sim2$ galaxies in \citet{Steidel2016} relative to the solar abundance pattern, given that $Z_*$ is primarily tracing Fe/H and $Z_{\textrm{ neb}}$ is tracing O/H. By analyzing a single composite rest-UV spectrum, \citet{Steidel2016} only probed average properties of their high-redshift galaxy sample. With single rest-UV and rest-optical composite spectra it is not possible to probe the average rest-UV spectral properties as a function of the location in the BPT diagram. In this paper we expand upon the important initial work of \citet{Steidel2016} by utilizing combined rest-UV and rest-optical spectra of 62 $z\sim2.3$ galaxies spanning a broad range of physical properties.  With this large sample, we investigate how the rest-UV spectral properties of the massive star population, including the inferred ionizing radiation field, vary for galaxies with different rest-optical emission-line properties in order to uncover the origin of differences between high-redshift and local galaxies in the BPT diagram.

The organization of this paper is as follows: Section 2 describes our observations, data reduction, and methods.  Section 3 presents the results of our analysis, and Section 4 provides a summary and discussion of our key results. Throughout this paper we assume a cosmology with $\Omega_m = 0.3$, $\Omega_{\Lambda}=0.7$, $H_0=70 \textrm{ km\ }s^{-1}\ \textrm{Mpc}^{-1}$, and adopt solar abundances from \citet{Asplund2009} (i.e., $Z_{\odot}=0.014$).

\section{Methods and Observations}

\subsection{Rest-Optical Spectra and the MOSDEF survey}
\begin{figure*}
    \centering
    \includegraphics[width=0.9\textwidth]{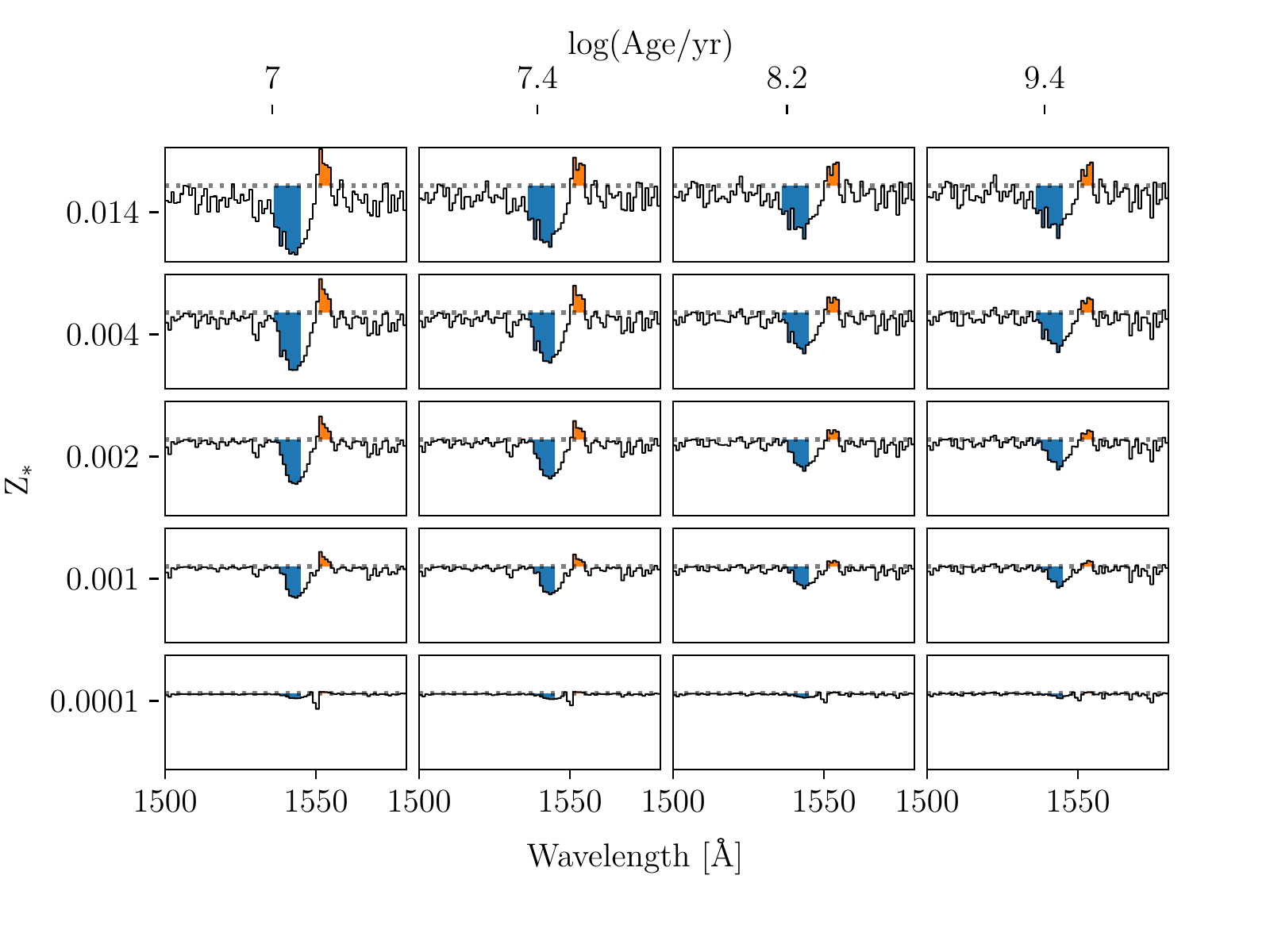} 
    \caption{Zoomed-in regions around the \CIV profile for several of the BPASS+Cloudy models of different stellar metallicities and ages used in our analysis.  The blueshifted wing of the broad wind absorption, and the redshifted wind emission feature of the \CIV profile are highlighted in blue and orange respectively, showing how the overall \CIV profile varies with age and metallicity.  Both the absorption (blue) and emission (orange) increase with strength toward higher stellar metallicity and younger age.  Both of these shaded regions are outside of the range where interstellar absorption is important and which we mask for our stellar population fitting analysis.}
    \label{fig:bptmodelwindlines}
\end{figure*}

Our analysis utilizes rest-optical spectroscopy of $z\sim2.3$ galaxies from the MOSDEF survey \citep{Kriek2015}. The MOSDEF survey consists of moderate resolution ($\textrm{R}\sim3500$) near-infrared spectra of $\sim1500$ $H$-band selected galaxies observed over 48.5 nights during 2012--2016 and targeted to lie within three distinct redshift intervals ($1.37 \le z \le 1.70$, $2.09 \le z \le 2.61$, and $2.95 \le z \le 3.80 $) near the epoch of peak star formation $(1.4\le z \le 3.8)$ using the Multi-Object Spectrometer for Infra-Red Exploration \citep[MOSFIRE;][]{McLean2012}. The actual redshift intervals are slightly different from our initial target ranges, based on the scatter between photometric and spectoscopic redshifts, and we redefine them as $1.40 \le z \le 1.90$, $1.90 \le z \le 2.65$, and $2.95 \le z \le 3.80 $. Additionally, the MOSDEF survey targeted galaxies in the \textit{Hubble Space Telescope} extragalactic legacy fields in regions covered by the CANDELS \citep{Grogin2011} and 3D-HST \citep{Momcheva2016} surveys, which have assembled extensive ancillary multi-wavelength datasets. MOSDEF spectra were used to measure fluxes and redshifts of all rest-optical emission lines detected within the Y, J, H, and K bands, the strongest of which are: [OII]$\lambda3727$, H$\beta$, [OIII]$\lambda \lambda 4959,5007$, H$\alpha$, [NII]$\lambda6584$, and [SII]$\lambda \lambda 6717,6731$.  

\subsection{LRIS Observations and Data}
In order to characterize how galaxy properties vary across the BPT diagram, we selected a subset of MOSDEF galaxies for rest-UV spectroscopic followup based on the following criterion. We prioritized selecting galaxies drawn from the MOSDEF survey for which all four BPT emission lines (H$\beta$, [OIII], H$\alpha$, [NII]) were detected with $\ge3\sigma$. Next highest priority was given to objects where H$\alpha$, H$\beta$, and [OIII] were detected, and an upper limit on [NII] was available.  Finally, in order of decreasing priority, the remaining targets were selected based on: availability of spectroscopic redshift measurement from MOSDEF (with higher priorities given to those objects at $1.90 \le z \le 2.65$ than those at $1.40 \le z \le 1.90$ or $2.95 \le z \le 3.80 $), objects observed as part of the MOSDEF survey without successful redshift measurements, and objects not observed on MOSDEF masks but contained within the 3D-HST survey catalog \citep{Momcheva2016} and lying within the MOSDEF target photometric redshift and apparent magnitude range.  These targets comprise $\sim 260$ observed galaxies with redshifts $1.4\le z \le 3.8$\footnote{Of the 260 observed galaxies, 214 galaxies had a redshift from the MOSDEF survey, with 32, 162, and 20 in the redshift intervals $1.40 \le z \le 1.90$, $1.90 \le z \le 2.65$, and $2.95 \le z \le 3.80 $ respectively.  The remaining 46 galaxies had either a spectroscopic redshift prior to the MOSDEF survey, or a photometric redshift, with 9, 31, and 6 in the redshift intervals $1.40 \le z \le 1.90$, $1.90 \le z \le 2.65$, and $2.95 \le z \le 3.80 $ respectively.}, which is large and diverse enough to create bins across multiple galaxy properties (e.g., location in the BPT diagram, stellar mass, SFR). For this analysis, we do not include the small fraction of objects identified as AGN based on their X-ray and rest-IR properties. Figure~\ref{fig:bptsample} displays the redshift histogram and distributions of $\textrm{ H}\alpha$-based SFR and $M_*$ derived from SED fitting \citep{Kriek2015} of the objects in our sample. A more detailed description of our method for SED fitting is described in Section~\ref{sec:results}.

A detailed description of the LRIS data acquisition and data reduction procedures will be presented elsewhere, however a brief summary is provided here. The data were obtained using the Low-Resolution Imaging Spectrograph \citep[LRIS;][]{Oke1995} during five observing runs totalling ten nights between January 2017 and June 2018. We observed 9 multi-object slit masks with $1\secpoint2$ slits in the COSMOS, AEGIS, GOODS-S, and GOODS-N fields targeting 259 distinct galaxies. We used the d500 dichroic, the 400 lines mm$^{-1}$ grism blazed at $3400\textrm{\angstrom}$ on the blue side, and the 600 lines mm$^{-1}$ grating blazed at $5000\textrm{\angstrom}$ on the red side.  This setup provided continuous wavelength coverage from the atmospheric cut-off at $3100\textrm{ \angstrom}$ up to a typical red wavelength limit of 7650$\textrm{ \angstrom}$. The blue side yielded a resolution of $R\sim800$, and the red side yielded a resolution of $R\sim1300$. The median exposure time was 7.5 hours, but ranged from 6--11 hours on different masks.  One night was lost completely due to weather. On 6/9 of the remaining nights the conditions were clear, and on 3/9 of the remaining nights there were some clouds, although we collected data on all three of those nights.  The seeing ranged from $0\secpoint6$ to $1\secpoint2$ with typical values of $0\secpoint8$. Details of the observations are listed in Table ~\ref{table:observations}.

\begin{table*}
\begin{center}
\begin{tabular}{rrrrrrr}
\toprule
Field & Mask Name &R.A. & decl.  & $t_{\textrm{ exp}}^{\textrm{ Blue}}$ [s] & $t_{\textrm{ exp}}^{\textrm{ Red}}$ [s] & $N_{\textrm{ objects}}$\\
\midrule
 COSMOS &$\textrm{ co\_l1}$ & 10:00:22.142 & $+$02:14:25.623  & $25200$ & $24080$ & 33 \\
 COSMOS &$\textrm{ co\_l2}$ & 10:00:22.886 & $+$02:24:45.096  & $24300$  & $22716$ & 31\\
 COSMOS &$\textrm{ co\_l5}$ & 10:00:29.608 & $+$02:14:33.037  & $21492$  & $20736$ & 27\\
 COSMOS &$\textrm{ co\_l6}$ & 10:00:39.965 & $+$02:17:28.409  & $25020$  & $24264$ & 26\\
 GOODS-S &$\textrm{ gs\_l1}$ & 03:32:23.178 & $-$27:43:08.900  & $39312$  & $38664$ & 30\\
 GOODS-N &$\textrm{ gn\_l1}$ & 12:37:13.178 & $+$62:15:09.647  & $27000$  & $22968$ & 30\\
 GOODS-N &$\textrm{ gn\_l3}$ & 12:36:54.841 & $+$62:15:32.920  & $32400$  & $31500$ & 27\\
 AEGIS &$\textrm{ ae\_l1}$ & 14:19:14.858 & $+$52:48:02.128  & $28188$  & $26964$ & 31\\
 AEGIS &$\textrm{ ae\_l3}$ & 14:19:35.219 & $+$52:54:52.570  & $34056$  & $33120$ & 25\\
 \bottomrule
 \end{tabular}
 \end{center}
 \caption{Summary of LRIS observations.}

\label{table:observations}
\end{table*}

\begin{figure*}
    \centering
    \includegraphics[width=1.0\linewidth]{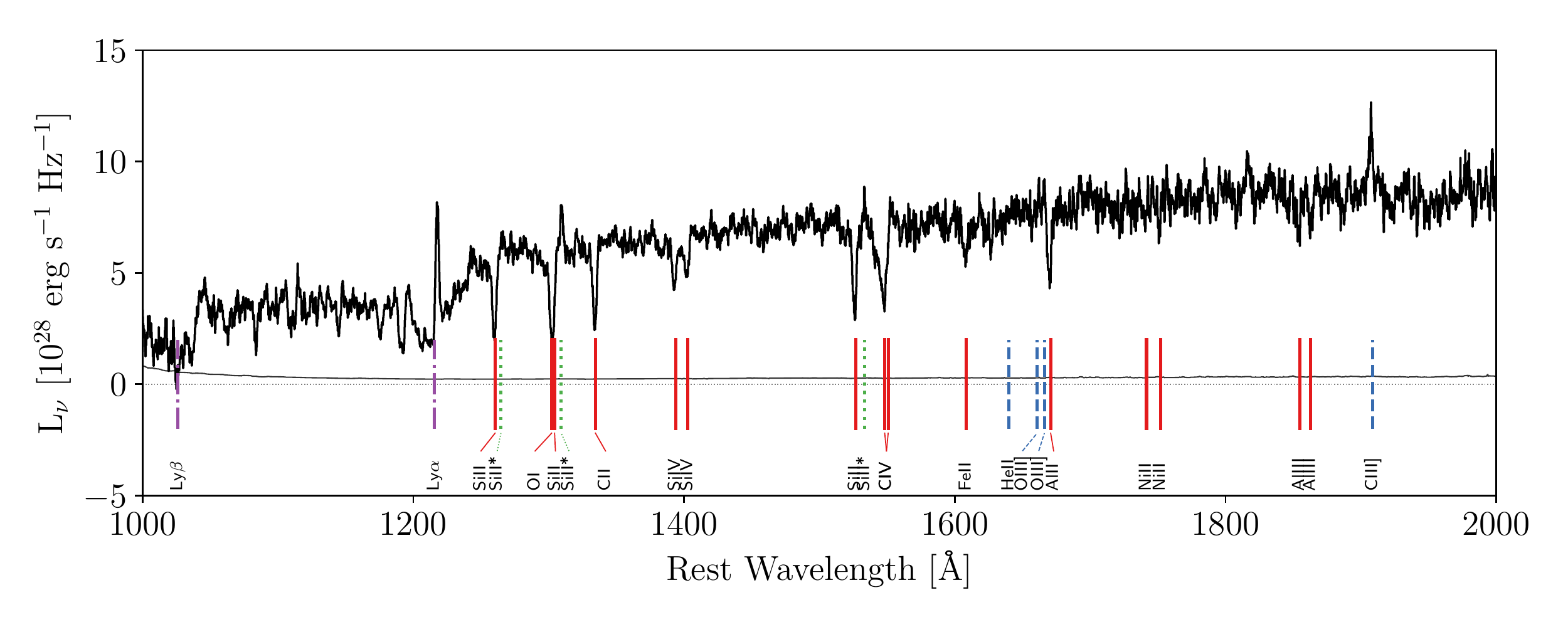}
    \caption{Stacked spectrum composed of all galaxies with a redshift measured from LRIS and in the interval $2.09\le z \le 2.55$, totalling 62 galaxies. The magnitude of the $1\sigma$ error spectrum is depicted by the thin black line. Spectral features are identified using the same labeling scheme as in Figure~\ref{fig:bptindividual}, with the addition of the Lyman series marked by dashed-dotted purple lines.}
    \label{fig:bptfullstack}
\end{figure*}

We reduced the data from the LRIS red and blue detectors using custom \texttt{iraf}, \texttt{idl}, and \texttt{python} scripts.  We first fit polynomials to the traces of each slit edge, and rectified each slit accordingly, straightening the slit-edge traces. For blue-side images, we then flat fielded each frame using twilight sky flats, and dome flats for the red side. We cut out the slitlet for each object in all flat-fielded exposures. Following this step we used slightly different methods to reduce the red- and blue-side images. For each object, the blue-side slitlets were first cleaned of cosmic rays.  Then, slitlets from each individual blue frame were background subtracted, registered and combined to create a stacked two-dimensional spectrum. We then performed a second-pass background subtraction on the stacked two-dimensional spectrum of each object while excluding the traces of objects in the slits in order to avoid over-subtraction of the background \citep{Shapley2006}. For the red-side images, we first background subtracted the individual frames, and cut out the slitlet for each object in all images. These individual slitlets were then registered and median combined using minmax rejection to remove cosmic rays, which more significantly contaminate the red-side slitlets. We then used the stacked two-dimensional spectra to measure the traces of objects in each slit.  The abundance of sky lines in the red-side images prevented us from achieving an accurate second-pass background subtraction on the stacked two-dimensional spectra. Therefore, we masked out the spectral traces in the individual red side slitlets, and recalculated the background subtraction on the individual slitlets. These individual, background subtracted slitlets were re-registered and median combined with rejection again to create the final stacked image.  

\begin{figure}
    \centering
    \includegraphics[width=\linewidth]{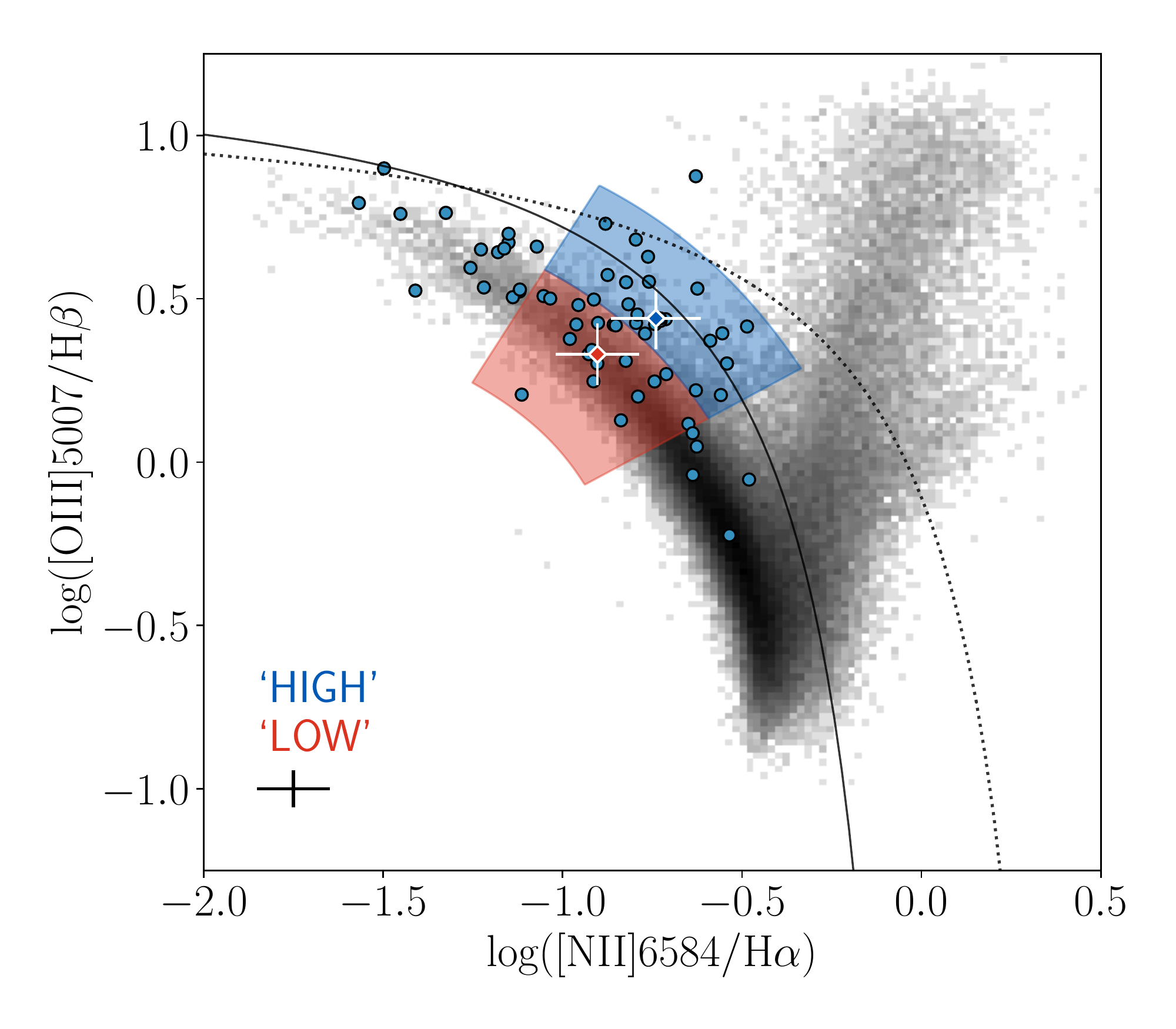}
    \caption{[O III]/H$\beta$ vs. [N II]/H$\alpha$ BPT diagram for $z\sim2.3$ galaxies in our sample (blue points) as well as local SDSS galaxies \citep[grey;][]{Abazajian2009}. We split the region of the BPT diagram most densely covered by our sample into two bins, one consisting of galaxies along the locus of $z=0$ galaxies (red shaded region; \textit{low} sample), and one bin of galaxies toward higher [O III]/H$\beta$ and [N II]/H$\alpha$ (blue shaded region;  \textit{high} sample). The median [OIII]/H$\beta$ vs. [N II]/H$\alpha$ values of galaxies in each bin are depicted as diamond-shaped symbols, with values of ([N II]/H$\alpha$, [OIII]/H$\beta$)=($-0.90\pm0.12$, $0.33\pm0.09$) for the \textit{low} stack, and ($-0.74\pm0.13$, $0.44\pm0.09$) for the \textit{high} stack.  For reference, the `maximum starburst' model of \citet{Kewley2001} (dotted curve) and star-formation/AGN boundary from \citet{Kauffmann2003} (solid curve) are plotted. A median error bar for the $z\sim2.3$ sample is shown in the bottom left.}
    \label{fig:bptregions}
\end{figure}

\begin{figure*}
    \centering
    \includegraphics[width=0.9\textwidth]{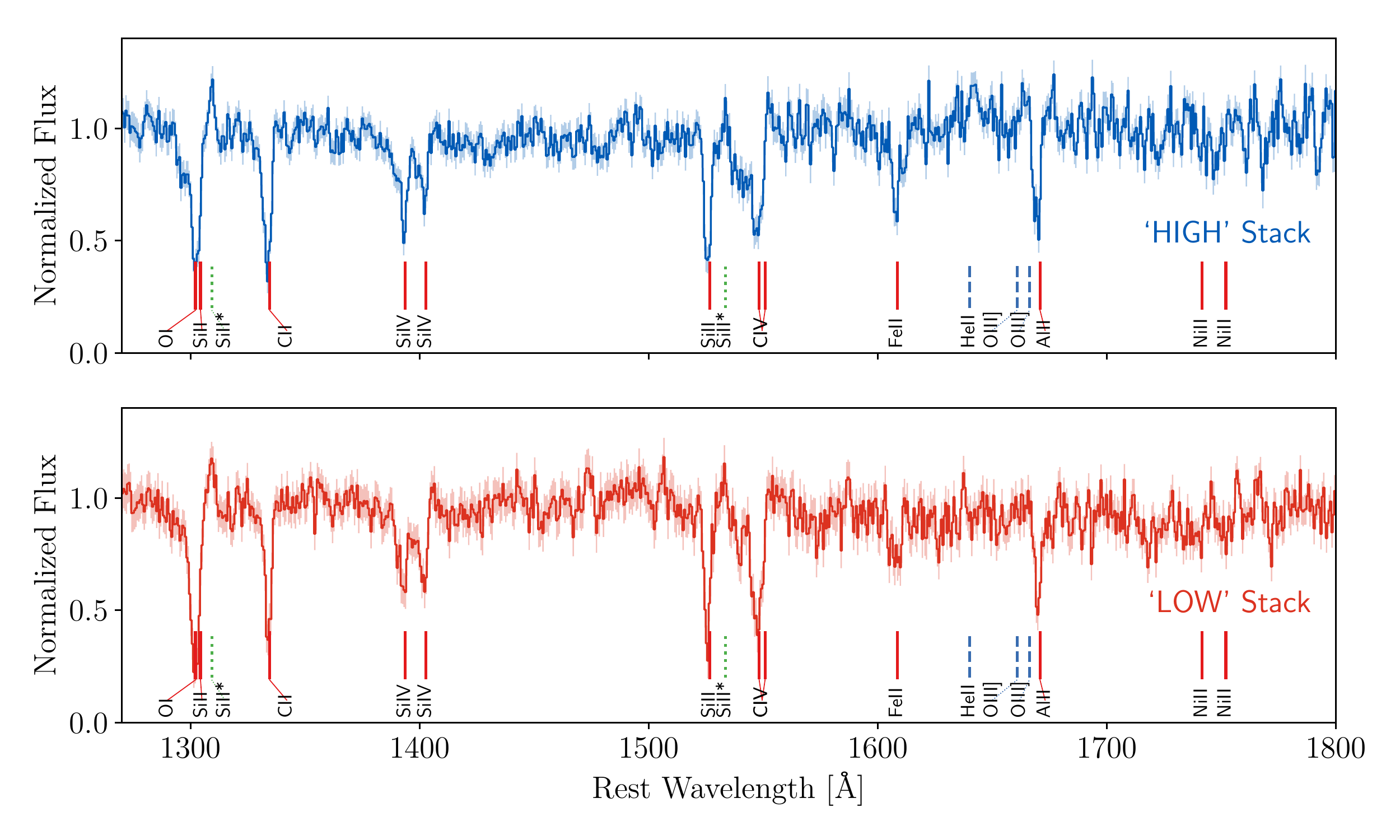}
    \caption{Top: A composite rest-UV spectrum of the 22 galaxies comprising our \textit{high} stack.  Bottom: A composite rest-UV spectrum of the 19 galaxies comprising our \textit{low} stack. Spectral features are labeled as in Figure~\ref{fig:bptindividual}.  Figure~\ref{fig:bptobswindlines} shows a zoomed-in comparison of the \CIV and \HeII lines. The $1\sigma$ composite error spectrum is indicated by the shaded region surrounding each spectrum.}
    \label{fig:bptdoubletstack}
\end{figure*}

Following these steps, we extracted and wavelength calibrated the blue and red side 1D spectrum of each object. The wavelength solution was calculated by fitting a 4th-order polynomial to the red and blue arc lamp spectra, resulting in typical residuals of $\sim0.035\textrm{ \angstrom}$ and $\sim0.3\textrm{ \angstrom}$ for the red and blue side spectra, respectively. We repeated this reduction procedure a second time without background subtraction and measured the centroid of several known sky lines. We shifted the wavelength solution zeropoint so that the sky lines appear at their correct wavelength values, and found the median required shift had a magnitude of $\sim4\angstrom$ in either direction. To apply the flux calibration, we used a first pass calibration based on spectrophotometric standard star observations obtained through a long slit during each observing run.  We performed a final, absolute flux calibration for each galaxy by comparing 3D-HST photometric measurements with spectrophotometric measurements calculated from our spectra, and normalized our spectra so that our calculated magnitudes matched the 3D-HST values.  After this absolute calibration, we checked that the continuum levels of the red and blue side spectra matched on either side of the dichroic cut-off at $5000\textrm{ \angstrom}$. Figure~\ref{fig:bptindividual} shows some examples of reduced high-SNR continuum normalized spectra.  Several strong absorption features are commonly visible, including Si\textsc{ii}$\lambda 1260$, O\textsc{i}$\lambda1302$+Si\textsc{ii}$\lambda1304$, C\textsc{ii}$\lambda1334$, Si\textsc{iv}$\lambda \lambda 1393, 1402$, \CIV, and Al\textsc{ii}$\lambda1670$.

\subsection{Redshift Measurements}
We measured a redshift for each object based on the Ly$\alpha$ emission line, as well as low-ionization interstellar (LIS) absorption lines, namely,  Si\textsc{ii}$\lambda 1260$, O\textsc{i}$\lambda1302$+Si\textsc{ii}$\lambda1304$, C\textsc{ii}$\lambda1334$, Si\textsc{ii}$\lambda1526$, Fe\textsc{ii}$\lambda1608$, and Al\textsc{ii}$\lambda1670$, where available.  Due to the presence of galaxy-scale outflows, the Ly$\alpha$ emission and interstellar absorption lines are commonly Doppler shifted away from the systemic redshift, $z_{\textrm{sys}}$. Therefore, we defined two different redshift measurements, $z_{\textrm{ Ly}\alpha}$, and $z_{\textrm{ LIS}}$.  We used the systemic redshift measured from nebular emission lines as part of the MOSDEF survey, when available, as an initial guess for $z_{\textrm{ Ly}\alpha}$, and $z_{\textrm{ LIS}}$.  If no redshift was present for an object in the MOSDEF survey, we manually inspected the LRIS spectrum and measured the redshift based on any available features. This manually measured redshift was then used as an initial guess for our redshift measurement analysis.  We measured the centroid of each line by simultaneously fitting the local continuum and spectral line with a quadratic function and a single Gaussian respectively.  We restricted the amplitude of the Gaussian to be $\ge0$ for the Ly$\alpha$ emission line, and $\le0$ for the absorption lines.  We repeated this fitting process 100 times for each line, and with every iteration we perturbed the spectrum  by its corresponding error spectrum.  The average and standard deviation of the centroids from the 100 trials became the measured redshift and uncertainty for each spectral line. We manually inspected the fits to each line, and excluded that line if the fits were poor. We calculated the final $z_{\textrm{ LIS}}$ using the available interstellar absorption lines for each galaxy by giving priority to absorption lines that provide a more accurate measurement of the redshift.  The Si\textsc{ii}$\lambda 1260$, C\textsc{ii}$\lambda1334$, and Si\textsc{ii}$\lambda1526$ absorption lines provide the best options to use as a redshift measurement, as they are not contaminated by nearby features \citep{Shapley2003}.  We averaged any successful redshift measurement from these three lines to obtain $z_{\textrm{ LIS}}$ (162 objects).  If an object did not have a redshift measurement for any of these three lines, we defined $z_{\textrm{ LIS}}$ by using the Al\textsc{ii}$\lambda1670$ line (1 object).  If this line was also not available we used the blended O\textsc{i}$\lambda1302$+Si\textsc{ii}$\lambda1304$ line (6 objects).  We established relations between systemic redshifts from the MOSDEF survey and redshift measurements from the rest-UV spectrum to infer the systemic redshift for galaxies without MOSDEF measurements.  In particular, we set $z_{\textrm{ sys}}$ as:
\[ \begin{cases} 
      z_{\textrm{ sys}}=z_{\textrm{ LIS}}+32.0 (\frac{1+z_{\textrm{ LIS}}}{c}) & z_{\textrm{ LIS}}\textrm{ only} \\
      z_{\textrm{ sys}}=z_{\textrm{ LIS}}+89.0 (\frac{1+z_{\textrm{ LIS}}}{c}) & z_{\textrm{ LIS}}\textrm{ and} \ z_{\textrm{ Ly}\alpha} \\
      z_{\textrm{ sys}}=z_{\textrm{ Ly}\alpha}-153.0 (\frac{1+z_{\textrm{ Ly}\alpha}}{c}) & z_{\textrm{ Ly}\alpha}\textrm{ \ only};\ z_{\textrm{ Ly}\alpha}\le2.7 \\
      z_{\textrm{ sys}}=z_{\textrm{ Ly}\alpha}-317.0 (\frac{1+z_{\textrm{ Ly}\alpha}}{c}) & z_{\textrm{ Ly}\alpha}\textrm{ \ only};\ z_{\textrm{ Ly}\alpha}\ge2.7. \\
   \end{cases}
\]
Finally, we used the systemic redshifts to shift each spectrum into the rest-frame. Out of the total 260 objects in our sample, 214 had systemic redshifts measured from the MOSDEF survey, 22 utilized our relations between $z_{\textrm{ sys}}$ and $z_{\textrm{ Ly}\alpha}$ or $z_{\textrm{ LIS}}$, and for the remaining 24 objects we were not able to measure a redshift.

\subsection{The LRIS-BPT Sample}
The full MOSDEF-LRIS sample consists of $260$ galaxies across three distinct redshift intervals ($1.40 \le z \le 1.90$, $1.90 \le z \le 2.65$, and $2.95 \le z \le 3.80 $).  We define a subset of this sample, hereafter referred to as the LRIS-BPT sample, which is composed of galaxies in the central redshift range that have detections in the four primary BPT emission lines (H$\beta$, [OIII]$\lambda 5007$, H$\alpha$, [NII]$\lambda6584$) at the $\ge3\sigma$ level from the MOSDEF survey and  a redshift measured from the LRIS spectrum.  These criteria result in a sample of 62 galaxies that we define as ``the LRIS-BPT sample.'' Due to the requirement of detections in the four rest-optical emission lines listed above, all 62 galaxies in this sample have a directly measured systemic redshift. Figure~\ref{fig:bptfullstack} displays the median-combined composite spectrum of the 62 galaxies in the LRIS-BPT sample. 

We compared the population of galaxies in the LRIS-BPT sample with that of the full MOSDEF survey. Figure~\ref{fig:bptsample} displays the SFR calculated from dust-corrected Balmer lines vs. $\textrm{ M}_*$ for both galaxies in the LRIS-BPT sample and the full MOSDEF sample. The LRIS-BPT sample is characterized by a median SFR of $\log(\textrm{SFR}/(M_{\odot}/\textrm{ yr}))=1.53\pm0.44$, and a median stellar mass of $\log(\textrm{M}_*/\textrm{M}_{\odot})=10.02\pm0.52$. The median values are consistent with the properties of galaxies in the central redshift range ($1.90 \le z \le 2.65$) of the full MOSDEF survey, which has a median SFR of $\log(\textrm{ SFR}/(M_{\odot}/\textrm{yr}))=1.36\pm0.50$ and median mass of $\log(\textrm{ M}_*/\textrm{M}_{\odot})=9.93\pm0.60$. The similarity in median SFRs for the LRIS-BPT and total MOSDEF $z\sim2$ samples also holds when using SFRs based on SED fitting, instead of from dust-corrected Balmer lines \citep{Shivaei2016}. These comparisons suggest that our LRIS-BPT sample is an unbiased subset of the full $z\sim2$ MOSDEF sample.

\subsection{Stellar Population Models}
In order to determine the physical properties of the stars within our target galaxies we compared their observed spectra to a grid of stellar population models created with varying parameters.  We used the Binary Population And Spectral Synthesis (BPASS) v2.2.1 models \citep{Eldridge2017, Stanway2018} because, relative to other recent models, they more accurately incorporate many key processes in the evolution of massive stars, including the addition of binary stars, rotational mixing, and Quasi-Homogeneous Evolution (QHE), resulting in longer main sequence lifetimes.  We considered BPASS stellar population models with all available stellar metallicities ($10^{-5} \le \textrm{ Z}_* \le 0.04$), which primarily trace Fe/H \citep{Steidel2016, Strom2018}, and ages between $10^7$yr and $10^{9.8}$yr in steps of 0.4 dex.  The upper limit in age for this grid was chosen to include the age of the universe at the lowest redshift in our sample. We used the stellar population models that assume a \citet{Chabrier2003} IMF, and have a high-mass cutoff of $100M_{\odot}$.  By default, BPASS provides models of an instantaneous burst of star formation.  We constructed models assuming a constant star-formation history, by summing up the burst models, weighted by their ages.

\begin{figure*}
    \centering
    \includegraphics[width=\linewidth]{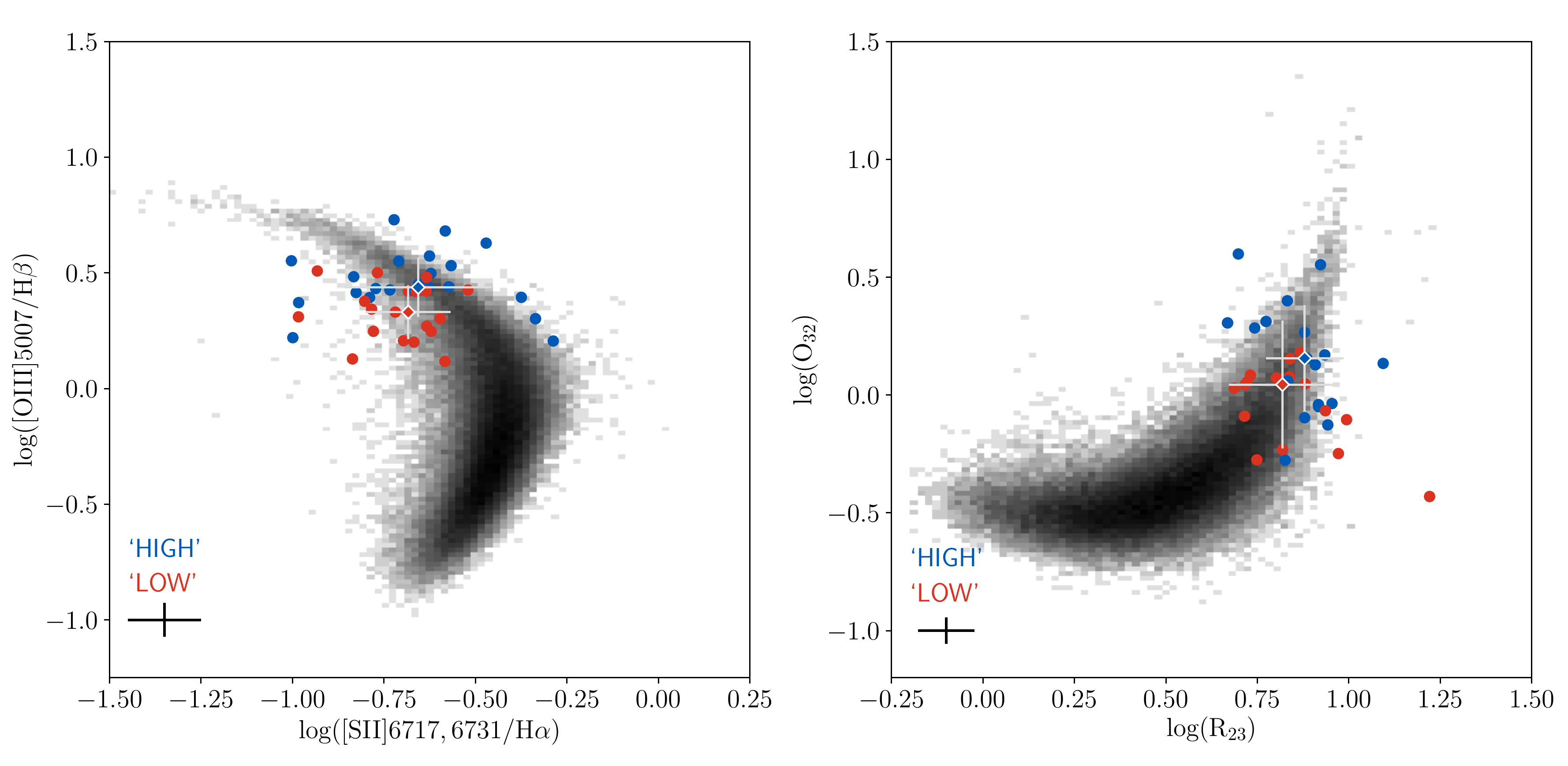}
    \caption{Left: [SII] BPT diagram. Galaxies in the \textit{high} and \textit{low} stacks are shown, respectively, using blue and red symbols. For comparison, the grey histogram shows the distribution of local SDSS galaxies. Galaxies in the \textit{high} stack are offset on average toward higher [OIII]$\lambda5007$/H$\beta$, however there is overlap between the two samples. A median errorbar is shown in the bottom left. Right: $\log(\textrm{O}_{32})$ vs. $\log(\textrm{R}_{23})$ diagram. Symbols are the same as in the left panel. Galaxies in the \textit{high} stack are offset on average toward higher $\log(\textrm{O}_{32})$ and $\log(\textrm{R}_{23})$, though there is overlap with the \textit{low} stack.}
    \label{fig:morebpt}
\end{figure*}

In order to accurately compare our models with our observed spectra we must include contributions from the nebular continuum.  To model the nebular continuum component of the UV spectrum we used the radiative transfer code Cloudy v17.01 \citep{Ferland2017}. For each individual BPASS stellar population of a given age and stellar metallicity, we ran a grid of Cloudy models with a range of nebular metallicities (i.e., gas-phase O/H) and ionization parameters.  Our Cloudy grids include a range of nebular metallicities of $-2.0 \le \log( Z_{\textrm{neb}}/Z_{\odot}) \le 0.4$ in 0.2 dex steps, and ionization parameters of $-4.0 \le \log(U) \le -1.0$ in 0.4 dex steps.  All models were run assuming an electron density typical of galaxies at this redshift of $n_e = 250\ \textrm{ cm}^{-3}$ \citep{Sanders2016, Strom2017}. We set the abundance of nitrogen in the models using the $\textrm{ log}(\textrm{ N/O})$ vs. $\textrm{ log}(\textrm{ O/H})$ relation from \citet{Pilyugin2012}: \vspace{0.4cm}

      \ $\log(\textrm{ N/O})=-1.493$ \\
     \vspace{.1cm}
      \indent \indent for $12+\log(\textrm{ O/H}) < 8.14$\\
           \vspace{.1cm}
     \indent $\log(\textrm{ N/O})=1.489 \times [12+\log(\textrm{ O/H})] - 13.613$ \\
          \vspace{.1cm}
      \indent \indent for $12+\log(\textrm{ O/H}) \ge 8.14$. 
 \vspace{0.1cm}

When using the stellar population models we added the contribution from the nebular continuum calculated assuming parameters typical of galaxies at this redshift \citep[$\log(U)=-2.5$, $\log( Z_{\textrm{neb}}/\textrm{ Z}_{\odot}) = -0.2$;][]{Sanders2016}. Adjusting these parameters does not affect the nebular continuum significantly enough to alter the results of our model fitting. Figure~\ref{fig:bptmodelwindlines} shows the differences in the \CIV profile for a subset of age and stellar metallicity models used in our analysis.  Two key features are highlighted in blue and red, both of which are located within regions of the \CIV profile that are not strongly affected by contamination from interstellar absorption.  Both of these features increase in strength towards higher stellar metallicity and younger ages.  While the strengths of these features do not necessarily represent a unique combination of age and stellar metallicity, this degeneracy is broken by considering the full rest-UV spectrum.

\subsection{Spectra fitting}
We fit the combined stellar population plus nebular continuum models to our observed spectra in order to determine which stellar population parameters produce a spectrum that most closely matches our observed spectra. We first continuum normalized the observed and model spectra. In fitting the continuum level, we only considered the rest-frame spectral region at $1270\textrm{ \angstrom} \le \lambda \le 2000\textrm{ \angstrom}$ to avoid the Ly$\alpha$ feature on the blue end, and a decrease in the quality of our spectra redwards of 2000$\textrm{ \angstrom}$.  To define an accurate continuum, we used spectral windows in regions of the spectrum relatively unaffected by stellar or nebular features, as defined by \citet{Rix2004}. We averaged the flux in each of the windows and fit a cubic spline through the windows to obtain the continuum level. 

The models that we used consist of stellar and nebular continuum components only, so we masked out regions of the spectrum that contain other features, such as interstellar absorption.  For this purpose, we adopted `Mask 1' from \citet{Steidel2016} in the wavelength range $1270\textrm{ \angstrom}-2000\textrm{ \angstrom}$. To determine the best-fit age and metallicity, we first interpolated the model onto the wavelength scale of our observed spectrum, and then calculated the $\chi^2$ for each model in our grid:
\begin{equation}
    \chi^2 = \sum_{i} \frac{(f_{\mathrm{spec},i} - f_{\mathrm{model},i})^2}{\sigma^2_i},
\end{equation}
where $f_{\mathrm{spec},i}$, $f_{\mathrm{model},i}$, and $\sigma^2_i$ are the individual pixel values of the masked, continuum-normalized observed spectrum, masked, continuum-normalized model spectrum, and variance in the spectrum, respectively. We did not smooth either the models or the observed spectra as their resolutions were comparable with values of $\sim1\angstrom$ in the rest-frame.  This sum was typically carried out over $\sim1000$ wavelength elements, and resulted in a $\chi^2$ surface in the $\log(\textrm{ Age/yr})$-$Z_*$ plane, which we interpolated using a 2D cubic spline and minimized to find the best-fit parameters.  To calculate the uncertainties in these parameters, we perturbed the spectrum and repeated this process 1000 times to produce a distribution of best-fit values.  We then defined the boundaries of the 1$\sigma$ confidence interval at the 16th and 84th percentiles of this distribution. 

In addition to fitting individual spectra, we applied our method to fit composite spectra.  To construct a composite spectrum, we first interpolated continuum-normalized individual spectra to a common wavelength grid with the sampling of the typical blue-side spectra (i.e., the lower resolution side), resulting in a typical sampling of $ \sim0.6 \textrm{ \angstrom}/ \textrm{pixel}$ in the rest frame.  We then median combined the interpolated spectra to produce the final composite spectrum.  We constructed the composite error spectrum using a bootstrap resampling method.  For a composite spectrum composed of a given number of galaxy spectra, we first selected an equal number of spectra from the composite sample with replacement.  We perturbed the selected spectra by their corresponding error spectra, and median combined them to create a composite spectrum.  This process was repeated 1000 times to assemble an array of composite spectra.  Finally, the composite error spectrum was determined as the standard deviation of the distribution of flux values of the perturbed composite spectra at each wavelength element.

\section{Results}
\label{sec:results}
To determine how galaxy properties vary across the BPT diagram, we create two stacks of galaxies with roughly comparable oxygen abundance based on their similar [NII]/H$\alpha$ values, but characterized by different rest-optical line ratios relative to the $z=0$ BPT excitation sequence. Figure~\ref{fig:bptregions} shows the regions we use to define our stacks. We label the stack of galaxies consistent with the $z=0$ BPT locus as the \textit{low} stack, and the stack of galaxies at higher [NII]/H$\alpha$ and [OIII]/H$\beta$ as the \textit{high} stack. The two stacks contain a majority of the galaxies in our LRIS-BPT sample, with the \textit{low} and \textit{high} stacks comprising 19 and 22 galaxies respectively.  Despite being composed of a large number of galaxies, each stack covers a small enough area on the BPT diagram to sample galaxies with similar emission line properties.  Figure~\ref{fig:bptdoubletstack} shows the stacked, continuum-normalized spectra of galaxies in the \textit{high} and \textit{low} stacks in blue (top) and red (bottom) respectively. For completeness, Figure~\ref{fig:morebpt} shows the positions of galaxies in our \textit{high} and \textit{low} stacks on the [OIII]$\lambda5007/$H$\beta$ vs. $[\textrm{ S}II]\lambda\lambda6717,6731/\textrm{ H}\alpha$ and $O_{32}$ vs. $R_{23}$ emission-line diagrams, where $O_{32}=[\textrm{ O}III]\lambda\lambda4959,5007/[\textrm{O}II]\lambda\lambda3726,3729$ and $R_{23}=([\textrm{ O}III]\lambda\lambda4959,5007+[\textrm{O}II]\lambda\lambda3726,3729)/H\beta$. On both of these additional BPT diagrams, the median positions of the two stacks are offset, however there is overlap between the samples.

\begin{figure}
    \centering
    \includegraphics[width=\linewidth]{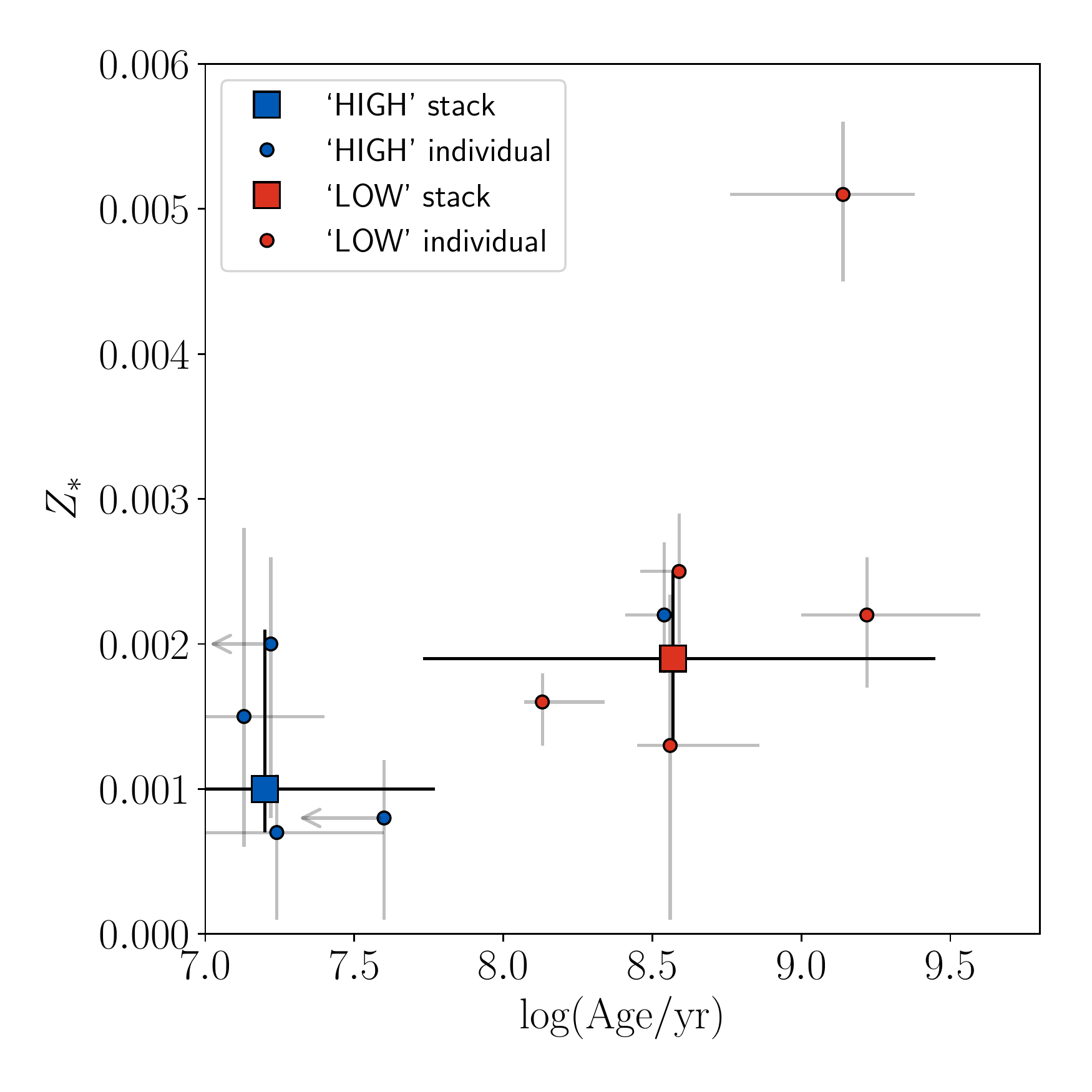}
    \caption{Best-fit $Z_*$ and $\log(\textrm{ Age/yr})$ for our two stacked spectra, and the five highest continuum SNR individual spectra from each stack, ranging from $4.5 \le \textrm{ SNR/pixel} \le 12$, measured in the wavelength range, $1425 \le \lambda \le 1500$. The remaining galaxies do not have high enough SNR ($\textrm{ SNR/pixel} \lesssim 4$) for our fitting procedure to produce reliable results without stacking. The large square points show results from the two stacks, and the small individual points are for individual galaxy measurements.  Points corresponding to the \textit{high} (22 galaxies) stack are indicated in blue, those from the \textit{low} (19 galaxies) stack are colored red.  The results from fitting individual galaxy spectra are predominantly consistent with the stacked results, however one galaxy from the \textit{high} stack has an age older than the stack.  Two galaxies from the \textit{high} stack had $1\sigma$ uncertainties at the edge of our grid, so they are represented as upper limits here.  The \textit{high} sample is characterized by a younger age and lower stellar metallicity compared to the \textit{low} stack.}
    \label{fig:bptfitting}
\end{figure}

In order to estimate the average physical properties of galaxies in our two stacks, we fit models to our stacked spectra using the procedure described above. To measure uncertainties in these properties, we repeated the fitting process 1000 times, during each of which we recreated the stack using galaxy spectra randomly chosen from the original stack with replacement and perturbed by their corresponding error spectrum.  Figure~\ref{fig:bptfitting} shows the best-fit stellar parameters that we determined for our two stacks.  Also shown are the results from applying our fitting procedure to the five individual galaxies with the highest SNR spectra in each bin.  We find a stellar metallicity of $Z_*=0.0010^{+0.0011}_{-0.0003}$ for the \textit{high} stack, and a stellar metallicity of $Z_*=0.0019^{+0.0006}_{-0.0006}$ for the \textit{low} stack. Both of these metallicities are consistent with each other within $1\sigma$. We find a best-fit stellar age for the \textit{low} stack of $\log(\textrm{ Age/yr})=8.57^{+0.88}_{-0.84}$. At this age, the number of O-stars, and therefore the FUV spectrum, has largely equilibrated in a stellar population with a constant star-formation history, which results in the large error bars \citep{Eldridge2012}. We find $\log(\textrm{ Age/yr})=7.20^{+0.57}_{-0.20}$ for the \textit{high} stack. This result suggests that the galaxies consistent with the \textit{high} stack typically have younger stellar populations compared to those in the \textit{low} stack. 

We check the properties for the galaxies in each stack estimated by comparing their broadband SEDs to stellar population synthesis models. Briefly, this analysis uses the fitting code FAST \citep{Kriek2009} to fit stellar population models from \citet{Conroy2009}, assuming a \citet{Chabrier2003} IMF and the \citet{Calzetti2000} dust reddening curve. The models also assume a ``delayed-$\tau$'' star-formation history of the form: $\textrm{ SFR}\propto t\ exp(-t/\tau)$, where $t$ is the time since the onset of star formation, and $\tau$ is the characteristic star formation timescale. For a full description of the SED fitting procedure see \citet{Kriek2015}. Based on the SED fitting, we find median stellar masses of $\log(\textrm{ M}/M_{\odot})=10.05\pm0.43$ and $\log(\textrm{ M}/M_{\odot})=10.12\pm0.32$ for galaxies in the \textit{high} and \textit{low} stacks respectively.  Also, we find median SFR of $\log(\textrm{ SFR}_{\textrm{ SED}}/(M_{\odot}/yr))=1.33\pm0.42$ and $\log(\textrm{ SFR}_{\textrm{ SED}}/(M_{\odot}/yr))=1.38\pm0.34$ for the galaxies in the \textit{high} and \textit{low} stacks respectively. Therefore, both stacks comprise galaxies that are well matched in SFR and $M_*$. Additionally, the median SED-based age for galaxies in the \textit{high} stack ($\log(\textrm{ Age/yr})=8.5\pm 0.4$) is younger than the median SED-based age in the \textit{low} stack ($\log(\textrm{ Age/yr})=8.6\pm 0.3$). This result from the broadband SED fitting agrees qualitatively with the younger age we find for the \textit{high} stack based on the full rest-UV fitting. However, the SED-based ages for the two stacks are not significantly different. The differences between the ages inferred from the rest-UV spectra, and those reported from SED fitting likely arise for a couple of reasons. First, the rest-UV fitting only accounts for light from the most massive stars, while the SED-based results also include information from longer wavelengths. In addition, for the fitting in this work, we only consider a constant star-formation history, and the SED fitting employs a larger range of `delayed-$\tau$' star-formation histories of the form $\tau\times e^{-t/\tau}$, where both $t$ and $\tau$ are fitted parameters. Incorporating more complex star-formation histories into our rest-UV fitting will be the subject of a future work.

In addition, the results from fitting model spectra to the high-SNR individual galaxy spectra are largely consistent with the results from using the stacked spectra.  Four out of five individual galaxies from the \textit{high} stack that we fit had stellar properties (age, stellar metallicity) consistent with stacked spectrum results, two of which were upper limits on the age.  The remaining galaxy had a best fit age that was substantially older than the stack.  All five individual galaxies in the \textit{low} stack are consistent with an older population, and all but one object showed consistent metallicities with the stack.  The best-fit parameters of the stacked spectra have larger uncertainties compared to the individual spectra, which suggests that our bootstrap resampling method is capturing galaxy-to-galaxy variations of age and $Z_*$ in our sample.

\begin{figure*}
    \centering
    \includegraphics[width=\textwidth]{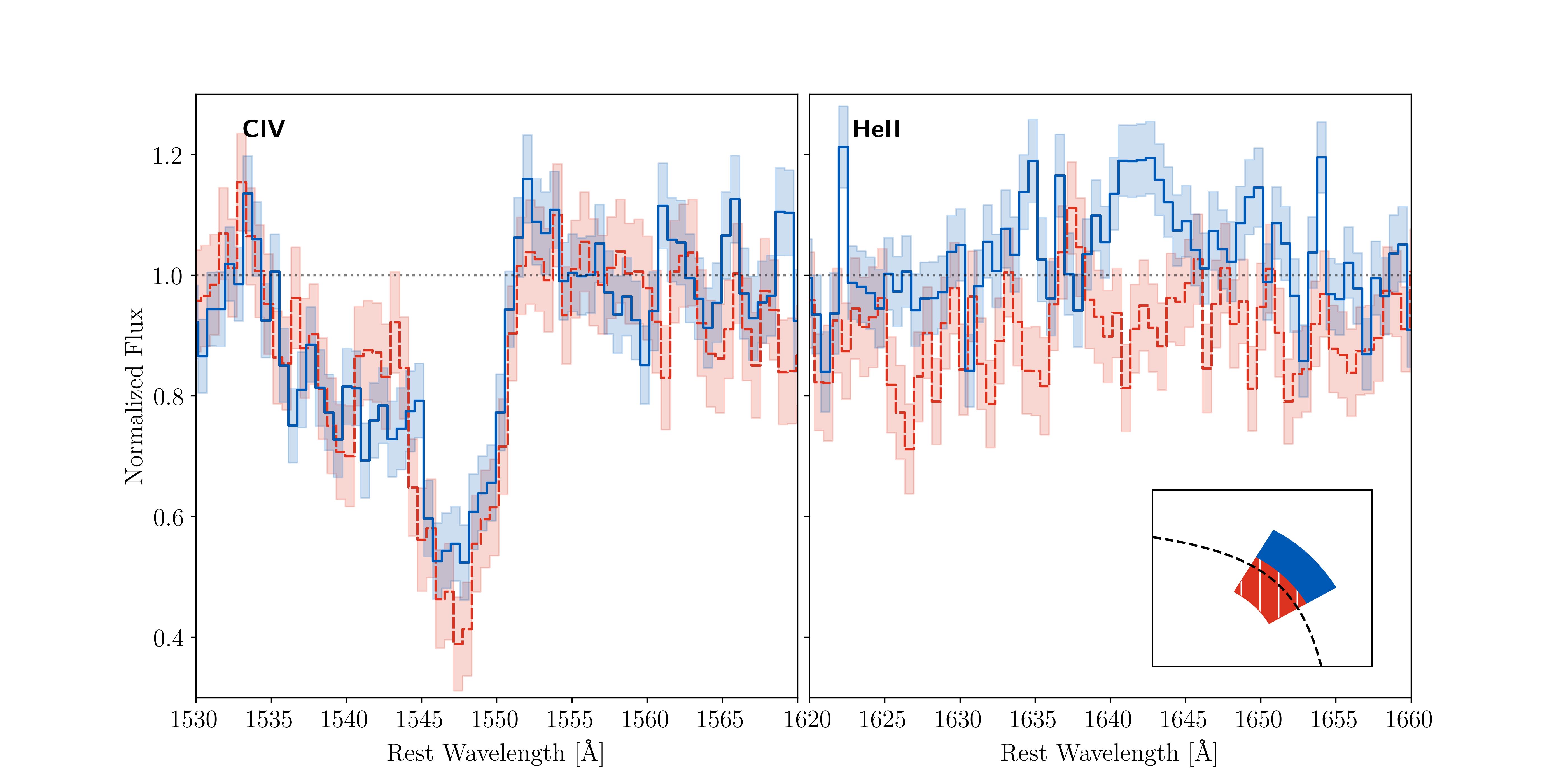}
    \caption{Zoomed-in regions around the \CIV (left) and \HeII (right) features for our \textit{high} (blue) and \textit{low} (red) stacked spectra.  The \textit{high} stacked spectrum has stronger \HeII and \CIV emission. Both features are signatures of massive stars, and are more prominent in younger populations. The $1\sigma$ composite error spectrum is depicted by the shaded region surrounding each spectrum. Figure~\ref{fig:bptmodelwindlines}  shows the age dependence of the \CIV line in the BPASS models.  }
    \label{fig:bptobswindlines}
\end{figure*}

In addition to our global rest-UV fitting procedure, which covers the full FUV spectrum at $1270\textrm{\angstrom} \le \lambda \le 2000\textrm{\angstrom}$, evidence for a difference in age between our two stacks is visible in the wind lines produced by massive stars: \CIV and \HeII.  Figure~\ref{fig:bptobswindlines} shows these features for both of our stacks. The \textit{high} stack has stronger \CIV emission ($1552\angstrom - 1555\angstrom$), as well as stronger stellar wind absorption ($1536\angstrom - 1545\angstrom$) when compared to the \textit{low} stack. This result is confirmed qualitatively by looking at the \CIV profiles produced by stellar population models, which predict stronger \CIV emission for younger stellar populations (Figure~\ref{fig:bptmodelwindlines}). In addition, the \textit{high} stack shows a significant \HeII emission line, whereas the \textit{low} stack has none visible.  Both of these features confirm the results of our fitting analysis suggesting that the stack of \textit{high} galaxies shows evidence for stellar youth.

\begin{figure}
    \centering
    \includegraphics[width=\linewidth]{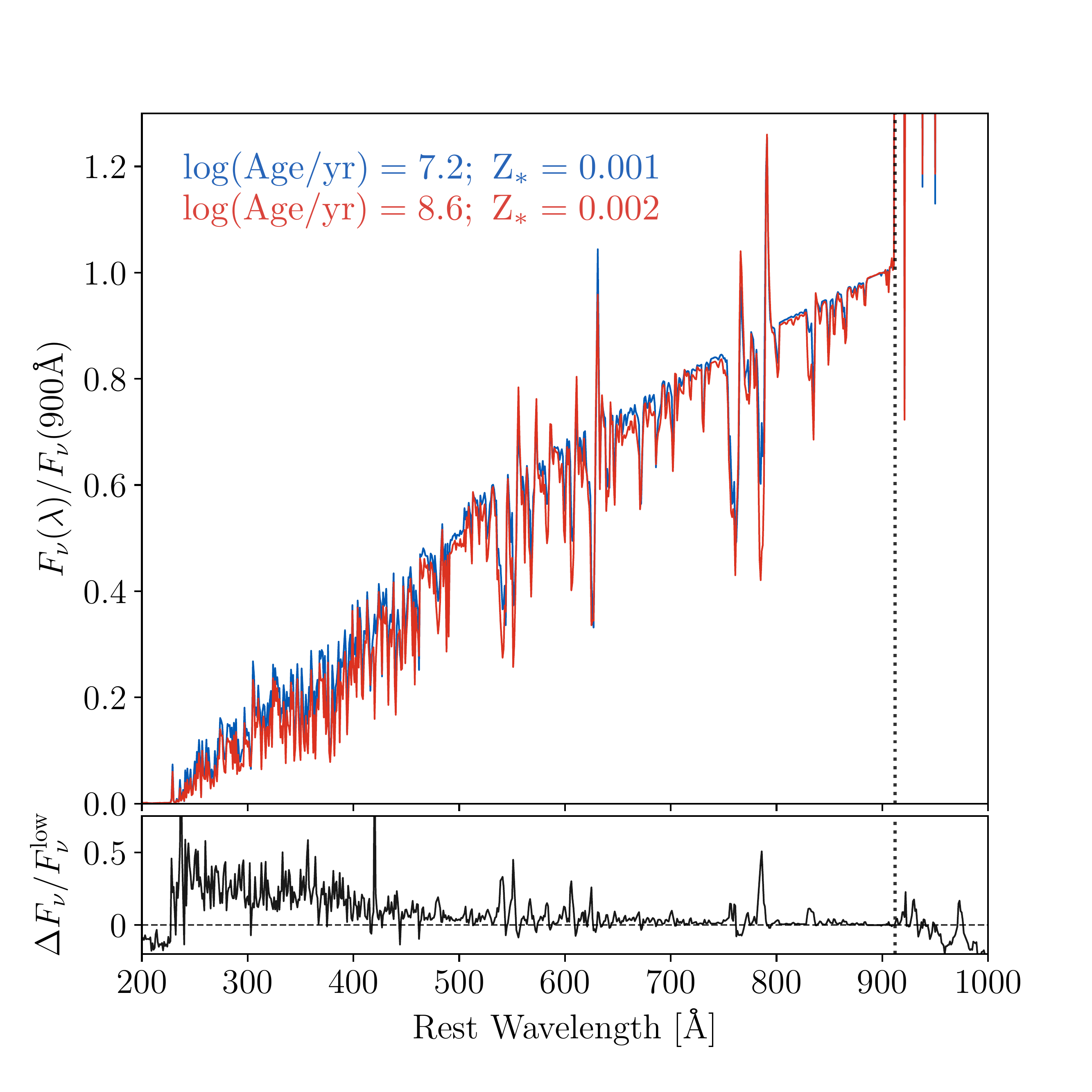}
    \caption{Top: Ionizing spectra of the best-fit stellar population models for the \textit{high} (blue) and \textit{low} (red) stacks.  The ionizing spectra are normalized at $900\textrm{\angstrom}$.  The normalized ionizing flux integrated in the range $200\textrm{ \angstrom} \le \lambda \le 912\textrm{ \angstrom}$ is $\sim7$\% higher in the \textit{high} stack compared to the \textit{low} stack. The vertical dotted line indicates the Lyman limit at $912\textrm{ \angstrom}$. Bottom: Fractional difference between the ionizing spectra of the best-fit stellar population models for the \textit{high} and \textit{low} stacks (i.e., $\Delta F_{\nu}/F_{\nu}^{\textrm{ low}}\equiv (F_{\nu}^{\textrm{ high}}-F_{\nu}^{\textrm{ low}})/F_{\nu}^{\textrm{ low}}$).}

    \label{fig:bptionizing}
\end{figure}

Using the best-fit stellar population parameters, we can examine the ionizing spectrum predicted by the BPASS models.  Figure~\ref{fig:bptionizing} shows the predicted ionizing spectrum for both the \textit{high} and \textit{low} stacks.  The most massive stars, which are responsible for producing the ionizing radiation, have lifetimes much shorter than the ages of most of our models.  Due to the assumed constant star-formation history in our models, the number of these massive stars equilibrates quickly ($\sim10$ Myr), and remains constant through most of our parameter space.  As a result, the ionizing spectrum is similar between the two best-fit models to our observed spectra, however the model corresponding to the \textit{high} stack has a harder ionizing spectrum due to its lower stellar metallicity. Specifically, the ionizing flux normalized at $900\textrm{\angstrom}$, and integrated over the range $200\textrm{ \angstrom} \le \lambda \le 912\textrm{ \angstrom}$, is $\sim7$\% higher in the \textit{high} stack compared to the \textit{low} stack.

\begin{figure*}
    \centering
    \includegraphics[width=0.8\textwidth]{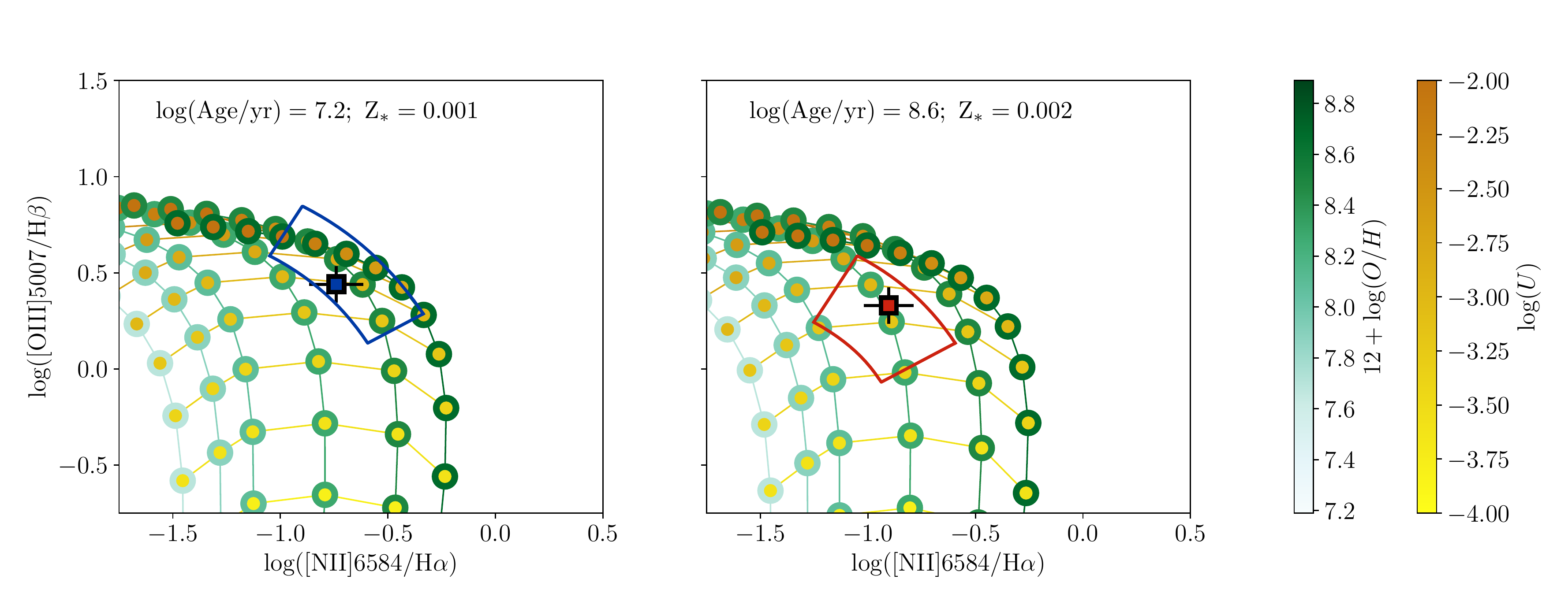}
    \caption{Predicted [OIII]/H$\beta$ and [NII]/H$\alpha$ emission-line ratios corresponding to the Cloudy+BPASS model grid of $12+\log(\textrm{ O/H})$ and $\log(U)$ for a given ionizing spectrum. The center of each point is color-coded by $\log(U)$, increasing from yellow to orange, while the border of each point is color-coded by $12+\log(\textrm{ O/H})$, increasing from light to dark green. The scale for each parameter is indicated by the color bars to the right of the panels. The median value and uncertainty in the observed [OIII]/H$\beta$ and [NII]/H$\alpha$ for galaxies in each bin are marked by the square points in each panel. The blue and red solid lines outline the regions of the BPT diagram we used to define our \textit{high} and \textit{low} stacks respectively. Left: Grid of line ratios assuming an ionizing spectrum corresponding to the best-fit stellar population for the high stack ($\log(\textrm{ Age/yr}) = 7.2,\ Z_*=0.001$). Right: Line ratios assuming an ionizing spectrum inferred from the best-fit stellar population model for the low stack ($\log(\textrm{ Age/yr}) = 8.6 \textrm{,\ Z}_*=0.002$).}
    \label{fig:bptnebulargrid}
\end{figure*}

\begin{figure*}
    \centering
    \includegraphics[width=0.8\textwidth]{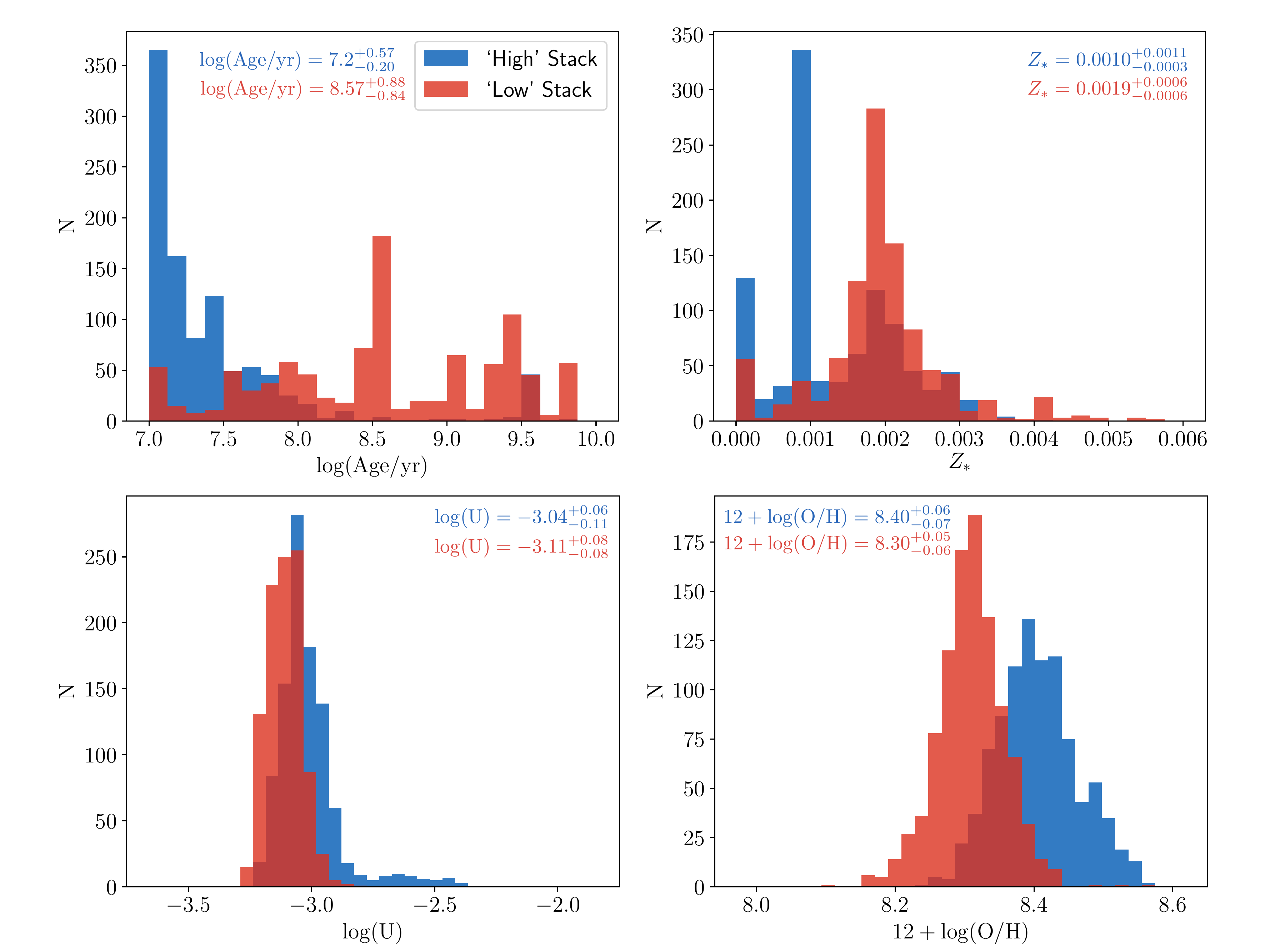}
    \caption{Distributions of best-fit age, stellar metallicity ($Z_*$), ionization parameter ($U$), and nebular metallicity (O/H) for our two stacks. The parameter distributions of age and stellar metallicity are produced by finding the minimum $\chi^2$ from fitting the grid of Cloudy+BPASS stellar population models to a bootstrap resampled composite spectrum 1000 times.  The parameter distributions of ionization parameter and nebular metallicity for each galaxy sample are produced by perturbing the sample median [O III]/H$\beta$ and [N II]/H$\alpha$ values by their uncertainties 1000 times and comparing the perturbed values to the inferred line ratios from our grid of Cloudy+BPASS models. All panels display parameter distributions for the \textit{high} and \textit{low} stacks in blue and red, respectively.  Top Left: Distribution of best-fit ages. We find the \textit{high} stack is younger ($\log(\textrm{ Age/yr})=7.20^{+0.57}_{-0.20}$) compared to the \textit{low} stack ($\log(\textrm{ Age/yr})=8.57^{+0.88}_{-0.84}$). Top Right: Distribution of best-fit stellar metallicities. We find that the \textit{high} stack has an overall lower stellar metallicity ($Z_*=0.0010^{+0.0011}_{-0.0003}$) compared to the \textit{low} stack ($Z_*=0.0019^{+0.0006}_{-0.0006}$). Bottom Left: Distributions of best-fit ionization parameters. We obtain values of $\log(U)=-3.04^{+0.06}_{-0.11}$ for the \textit{high} stack, and $\log(U)=-3.11^{+0.08}_{-0.08}$ for the \textit{high} stack.  Bottom Right: Distributions of best-fit nebular metallicities.  We find best-fit values of $12+\log(\textrm{ O/H})=8.40^{+0.06}_{-0.07}$ for the \textit{high} stack, and $12+\log(\textrm{ O/H})=8.30^{+0.05}_{-0.06}$ for the \textit{low} stack.}
    \label{fig:bptnebularparameters}
\end{figure*}

Using the predicted ionizing spectrum from our fitting analysis, we infer the nebular line fluxes expected for a given set of nebular parameters using Cloudy. We place our grid of Cloudy models on the [OIII]/H$\beta$ vs. [NII]/H$\alpha$ BPT diagram for the best-fit stellar spectrum of each of our stacks (Figure~\ref{fig:bptnebulargrid}). We linearly interpolate the grid of [NII]/H$\alpha$ and [OIII]/H$\beta$ values produced by the Cloudy models to determine which $Z_{\textrm{ neb}}$ and $\log(U)$ best match the median observed line ratios of each stack.  To estimate the uncertainty, we perturb the median observed [NII]/H$\alpha$ and [OIII]/H$\beta$ of the stacks by their uncertainties, and repeat the process 1000 times to create a distribution of values. Figure~\ref{fig:bptnebularparameters} (bottom row) displays the distributions of nebular metallicity and ionization parameter obtained from this analysis. We find an ionization parameter of $\log(U)=-3.04^{+0.06}_{-0.11}$  and nebular metallicity of $12+\log(\textrm{ O/H})=8.40^{+0.06}_{-0.07}$  for the \textit{high} stack.  For the \textit{low} stack, we find an ionization parameter of $\log(U)=-3.11^{+0.08}_{-0.08}$ and nebular metallicity of $12+\log(\textrm{ O/H})=8.30^{+0.05}_{-0.06}$. While these differences are consistent to $\sim1\sigma$, and are small given the dynamic range of ionization parameter in high-redshift star-forming galaxies, and systematic uncertainties in nebular metallicities, they have a measurable effect on the rest-optical emission ratios for the \textit{high} and \textit{low} stacks. We achieve similar results by instead fixing the ionizing spectrum for all galaxies in each stack, and inferring a distribution of nebular metallicities and ionization parameters of individual objects within the stack using the same method described above. Furthermore, we find the \textit{high} and \textit{low} stacks comprise samples with comparable electron density distributions, with median values of $n_e^{high}=350\pm161\textrm{ cm}^{-3}$ and $n_e^{low}=334\pm282\textrm{ cm}^{-3}$ respectively. Both medians are consistent with the value assumed in the Cloudy models ($n_e=250\textrm{ cm}^{-3}$), and a Kolmogorov-Smirnov test determines a $47\%$ probability that both samples are drawn from the same parent distribution. Table~\ref{table:physicalparameters} summarizes the best-fit physical parameters we find for the \textit{high} and \textit{low} stacks. 

\begin{table*}
\begin{center}
\renewcommand{\arraystretch}{1.4}
\begin{tabular}{rrrrr}
\toprule
  & $\log(\textrm{ Age/yr})$ &$Z_*$ & $\log(U)$  & $12+\log(\textrm{ O/H})$ \\
\midrule
 \textit{high} stack & $7.20^{+0.57}_{-0.20}$ & $0.0010^{+0.0011}_{-0.0003}$ & $-3.04^{+0.06}_{-0.11}$  & $8.40^{+0.06}_{-0.07}$  \\
 \textit{low} stack & $8.57^{+0.88}_{-0.84}$ & $0.0019^{+0.0006}_{-0.0006}$ & $-3.11^{+0.08}_{-0.08}$  & $8.30^{+0.05}_{-0.06}$  \\

 \bottomrule
 \end{tabular}
 \end{center}
 \caption{Best-fit physical parameters for the \textit{high} and \textit{low} stacks.}

\label{table:physicalparameters}
\end{table*}

\section{Discussion}
Sensitive multiplexed spectroscopic instruments on large telescopes have enabled the study of rest-optical spectra for statistical samples of galaxies at high-redshift. These studies have established an offset of high-redshift star-forming galaxies towards higher [O III]/H$\beta$ and [N II]/H$\alpha$ compared to local galaxies.  Several contributing factors have been proposed as the source of this offset, including varying abundance patterns, changes in the ionizing spectra, stellar and nebular metallicities, different ages of the stellar populations, and a different ionization parameter \citep{Shapley2005, Erb2006, Liu2008, Kewley2013}. One essential aspect of understanding these differences in high-redshift galaxies is a robust constraint on the ionizing spectrum produced by massive stars.  Rest-UV spectroscopy of star-forming galaxies traces the properties of the massive star populations, and (given a set of stellar population synthesis modeling assumptions) provides constraints on the ionizing radiation field. In turn, photoionization modelling enables us to connect the ionizing spectrum and massive star population, with rest-optical nebular line ratios including those in the BPT diagram. 

Currently, studies utilizing this combined rest-UV and rest-optical analysis have focused on average properties of the high-redshift population. By dividing our sample into two bins based on their location on the BPT diagram, we investigated how stellar population properties change as galaxies move away from the local BPT sequence. We found that the stack of galaxies above the local BPT sequence have younger ages, and lower stellar metallicities compared to galaxies along the local sequence at $z\sim2$.  Additionally, we find that galaxies above the local BPT sequence have harder ionizing spectra compared to their \textit{low} stack counterparts. In our models, which assume a constant star-formation history, the most massive star population equilibrates on timescales $\sim10$ Myr.  Therefore, the difference in ionizing spectrum is not due to the age difference between our two stacks, but instead the lower stellar metallicity (i.e., Fe/H) in those galaxies that are offset.

While the most notable difference between our \textit{low} and \textit{high} stacks is a factor of $\sim2$ lower stellar metallicity in the \textit{high} stack, our photoionization modelling reveals small differences in the additional nebular parameters $U$ and $Z_{\textrm{ neb}}$.  All three of these parameters contribute to the observed rest-optical emission line ratios of the \textit{high} and \textit{low} stacks.   Using photoionization modelling to measure $Z_{\textrm{ neb}}$ (nebular O/H), and using rest-UV spectral fitting to measure $Z_*$ (stellar Fe/H), we find that both stacks have super-solar O/Fe, with our \textit{low} and \textit{high} stacks having values of $3.04^{+0.95}_{-0.54}\textrm{ O/Fe}_{\odot}$ and $7.28^{+2.52}_{-2.82}\textrm{ O/Fe}_{\odot}$ respectively. While $\alpha$-enhancement has previously been presented as an explanation for the offset of $z\sim2$ galaxies in the BPT diagram, we stress that even galaxies that are entirely consistent with the local excitation sequence in the [OIII]$\lambda 5007$/H$\beta$ vs. [NII]$\lambda 6584$/H$\alpha$ diagram (i.e., the \textit{low} stack) appear to be $\alpha$-enhanced -- in contrast with local systems. Such differences must be considered in order to accurately model the properties of these galaxies and to infer gas-phase oxygen abundances based on strong emission-line ratios. Without accounting for these differences, models will produce nebular metallicities biased toward higher $12+\log(\textrm{ O/H})$. The O/Fe value for the \textit{high} stack is above the $\sim5.5 \times \textrm{ O/Fe}_{\odot}$ theoretical limit assuming a Salpeter IMF and high-mass cutoff of $50\ M_{\odot}$ \citep{Nomoto2006}, but is still consistent within $1\sigma$. However, the exact value of this theoretical limit is dependent on supernova yield models, which are not well constrained \citep{Kobayashi2006}. \citet{Kriek2016} found comparable $\alpha$-enhancement in a massive quiescent galaxy at $z=2.1$, reporting a $\textrm{ Mg/Fe}=3.9 \times \textrm{ Mg/Fe}_{\odot}$. 

The assumed nitrogen abundance at fixed O/H affects where photoionization model grids fall in the BPT diagram, such that increasing N/O increases [NII]/H$\alpha$ while keeping all other parameters fixed.  Consequently, if our assumed N/O-O/H relation does not hold for typical $z\sim2$ galaxies, then our inferred oxygen abundances will be systematically biased.  An underestimate in N/O leads to an overestimate of O/H, and vice versa. Therefore, the high $\alpha$-enhancement inferred in our offset galaxy stack could be due in part to differences in N/O at fixed O/H, perhaps due to the timescale of nitrogen enrichment in stellar populations \citep{Berg2019}.  However, in order for the \textit{high} and \textit{low} stacks to each have solar O/Fe, an enhancement of N/O by $\sim1$ dex and $\sim0.5$ dex respectively at fixed O/H would be required. Given the age of both stacks, and the timescale of Fe enrichment from Type Ia supernovae ($\sim1$ Gyr), the absence of $\alpha$-enhancement in either stack is unlikely. Another question is whether the difference in inferred $\alpha$-enhancement for the two stacks can be explained by different N/O vs. O/H relations. For O/Fe to match between the {\it high} and {\it low} stacks, we would need to assume an N/O higher by $\sim0.6$ dex for the {\it high} stack. For consistency at the $1\sigma$ level, the assumed N/O would need to be $\sim$0.2 dex higher for the {\it high} stack. Additionally, an O/Fe exceeding the theoretical limit of \citet{Nomoto2006} could be explained by a top-heavy IMF, or by increasing the high-mass cutoff of the stellar population.  Investigating these possible differences in stellar populations is an avenue for future analysis. 

To verify that our assumptions for the N/O ratio are reasonable, we compute the N/O ratio using the tracer, [NII]/[OII], for all objects in our stacks that have detections with $>3\sigma$ in both lines. We find that the high and low stacks are characterized by a median $[\textrm{ N}II]/[\textrm{ O}II] = -0.79\pm0.25$ and $[\textrm{ N}II]/[\textrm{ O}II] = -0.99\pm0.31$ respectively.  Based on the calibration of N/O as a function of [NII]/[OII] from \citet{Strom2018}, these line ratios correspond to a $\log(\textrm{ N/O})=-1.05\pm0.13$ for the high stack, and $\log(\textrm{ N/O})=-1.15\pm0.16$ for the low stack. Using the N/O to O/H relation from \citet{Pilyugin2012}, and the inferred nebular metallicity for out two stacks, we infer a nitrogen abundance of $\log(\textrm{ N/O})=-1.1$ for the high stack, and $\log(\textrm{ N/O})=-1.25$ for the low stack. These inferred values are both consistent with the nitrogen abundances computed based on [NII]/[OII], suggesting that our spectra are well described by the models.

We check the predicted $O_{32}$ distribution for the best-fit nebular metallicity and ionization parameter inferred from our models, and compare it to the observed $O_{32}$ distributions for our two stacks. We find that, on average, models for galaxies in the high stack have $O_{32}=0.16\pm0.22$ while models for galaxies in the low stack have $O_{32}=0.04\pm0.13$.  These values are in agreement with the distributions of observed $O_{32}$ measured from galaxies in our two stacks, for which we find $O_{32}=0.15\pm0.22$ for the high stack, and $O_{32}=0.03\pm0.16$ for the low stack. This agreement suggests that the best-fit models can self-consistently reproduce the observed $O_{32}$ line ratio.

An intriguing question is if the high-redshift galaxies that lie along the local sequence (i.e., the \textit{low} sample) be interpreted as descendants of the offset galaxies (i.e., the \textit{high} sample). Qualitatively, it is suggestive that this may be the case based on the age dependence of $\alpha$-enhancement seen in galactic bulge stars \citep{Matteucci2016}. However, chemical evolution models that incorporate realistic timescale differences between core collapse and Type Ia supernovae predict that significant evolution of O/Fe will only occur on timescales of $\sim1$ Gyr, assuming smooth star-formation histories \citep{Weinberg2017}, which is significantly longer than the age difference inferred between the {\it high} and {\it low} rest-UV composite spectra.  In contrast, in the models of \citet{Weinberg2017}, a sudden burst of star formation could temporarily boost O/Fe by $\sim0.3$ dex. Accordingly, galaxies in the {\it high} stack may show the evidence of recent bursts of star formation, and follow systematically different star-formation histories from those in the {\it low} stack. More detailed modelling will be required to see if this proposed explanation is applicable.

\section{Summary \& Conclusions}

We have obtained rest-UV spectra for a sample of 259 galaxies at $1.4\le z \le 3.8$ that were observed as part of the MOSDEF survey, enabling a combined analysis of rest-UV probes of massive stars and rest-optical probes of ionized gas.  Of these galaxies, 62 are at $z\sim 2.3$ ($2.09\le z \le 2.55$), and have all four BPT emission lines (H$\beta$, [OIII]$\lambda 5007$, H$\alpha$, [NII]$\lambda6584$) detected at $\ge 3 \sigma$. We constructed two composite rest-UV spectra of a subset of these 62 galaxy spectra based on their location on the BPT diagram.  We tested how galaxy properties, including the age, stellar metallicity, nebular metallicity, and ionization parameter vary for galaxies on and off the local sequence.  To derive these properties, we first fit a grid of Cloudy+BPASS stellar population synthesis models to constrain the age and stellar metallicity of the massive star population, therefore fixing the intrinsic ionizing spectrum. With the ionizing spectrum established, we then computed optical emission line flux ratios using Cloudy for a grid of nebular metallicities and ionization parameters.  Finally, we set the nebular metallicity and ionization parameter for our spectra based on the models that best reproduced the observed rest-optical emission line ratios. We summarize our main results and conclusions below.

(i) Using Cloudy+BPASS stellar population synthesis models we investigated how the age and stellar metallicity varies for high-redshift galaxies that lie on the local BPT sequence compared to those that are offset toward higher [OIII]$\lambda 5007$/H$\beta$ and [NII]$\lambda 6584$/H$\alpha$.  We found that the offset galaxies have younger ages ($\log(\textrm{ Age/yr})=7.20^{+0.57}_{-0.20}$) compared to the galaxies in our sample that lie on the local sequence ($\log(\textrm{ Age/yr})=8.57^{+0.88}_{-0.84}$).  Additionally, we found that the offset galaxies had overall lower stellar metallicities ($Z_*=0.0010^{+0.0011}_{-0.0003}$) compared to the non-offset galaxies ($Z_*=0.0019^{+0.0006}_{-0.0006}$).  These results are displayed in Figure~\ref{fig:bptfitting}.

(ii) We investigated how the ionizing spectrum of the best-fit stellar population synthesis models varies across the BPT diagram, and found that the galaxies that are offset from the local BPT sequence have a harder ionizing spectrum compared to those that are not offset (Figure~\ref{fig:bptionizing}). This difference is due to the lower stellar metallicity in the offset galaxies.  Inferred ages for both composites are old enough such that in constant star-formation models, the number of O-stars has reached an equilibrium, and the age of the population no longer has a significant effect on the ionizing spectrum.

(iii) Using the ionizing spectrum inferred for each stack from the rest-UV spectral fitting, we computed the resulting emission line fluxes for a grid of nebular metallicity, $Z_{\textrm{ neb}}$, and ionization parameter, $U$ (Figure~\ref{fig:bptnebulargrid}). Accordingly, our rest-UV spectral analysis enabled us to fix one of the input free parameters for photo-ionization modeling -- i.e., the form of the ionizing spectrum. We compared the resulting emission line flux ratios to the median observed ratios of our stacks from the MOSDEF survey in order to infer $Z_{\textrm{ neb}}$ and $U$ for our two galaxy stacks.  We found that the offset (\textit{high}) galaxies have an ionization parameter of $\log(U)=-3.04^{+0.06}_{-0.11}$ and the non-offset (\textit{low}) galaxies have an ionization parameter of ($\log(U)=-3.11^{+0.08}_{-0.08}$). In addition, the offset galaxy stack has a slightly higher nebular metallicity ($12+\log(\textrm{ O/H})=8.40^{+0.06}_{-0.07}$) compared to the non-offset galaxy stack ($12+\log(\textrm{ O/H})=8.30^{+0.05}_{-0.06}$).  The stellar and nebular metallicities we derived for our \textit{high} and \textit{low} stack imply that the galaxies that are offset from the local BPT relation are more $\alpha$-enhanced ($7.28^{+2.52}_{-2.82}\textrm{ O/Fe}_{\odot}$) compared to those on the local sequence ($3.04^{+0.95}_{-0.54}\textrm{ O/Fe}_{\odot}$). 

Understanding the observed differences between local and high-redshift galaxies in terms of their physical properties is required for a complete galaxy evolution model.  Thus far, these differences have mainly been probed in a sample-averaged sense, therefore variations across the high-redshift galaxy population cannot be determined.  By stacking our sample based on BPT location we observed which differences were enhanced in high-redshift galaxies that are most offset from the local sequence. We found that high-redshift galaxies had several factors contributing to the offset, namely that the most offset galaxies have younger ages, lower stellar metallicities, higher ionization parameters, and higher nebular oxygen abundances.  Notably, the offset galaxies are more $\alpha$-enhanced compared to high-redshift galaxies that lie along the local sequence.  Any photoionization modelling of $z\sim2.3$ galaxies that do not take these differences into account, instead using local properties, will yield biased results.  While $\alpha$-enhancement was found to be heightened in the most offset galaxies, some level of enhancement is present throughout the high-redshift sample--even those coincident with the local sample. Therefore, interpreting the agreement between the location local galaxies and some high-redshift galaxies (i.e., our \textit{low} sample) on the BPT diagram as a similarity of physical properties is an oversimplification.  While our method of inferring $Z_*$ from rest-UV spectral fitting, and $Z_{\textrm{ neb}}$ from photoionization modelling has not been applied to local galaxies, joint studies of the local stellar and gas-phase mass-metallicity relations suggest that $\sim L*$ star-forming galaxies in the local universe are not $\alpha$-enhanced \citep{Zahid2017}.

While we have refined the results of previous studies by measuring variations in high-redshift galaxy properties on and off the local sequence, a further refinement of composite spectra, or large numbers of high-SNR individual galaxies is still required. In addition, during this analysis we made several assumptions about the stellar populations of these galaxies, namely constant star-formation histories, and a single IMF.  Future investigations will need to examine more general star-formation histories and variations in the IMF in order to more accurately constrain galaxy properties at high redshift.

\section*{Acknowledgements}
We thank the anonymous referee for their helpful comments. We acknowledge support from NSF AAG grants AST1312780, 1312547, 1312764, and 1313171, grant AR13907 from the Space Telescope Science Institute, and grant NNX16AF54G from the NASA ADAP program. We also acknowledge a NASA contract supporting the ``WFIRST Extragalactic Potential Observations (EXPO) Science Investigation Team" (15-WFIRST15-0004), administered by GSFC. This work made use of v2.2.1 of the Binary Population and Spectral Synthesis (BPASS) models as described in Eldridge, Stanway et al. (2017) and Stanway \& Eldridge et al. (2018). We wish to extend special thanks to those of Hawaiian ancestry on whose sacred mountain we are privileged to be guests. Without their generous hospitality, most of the observations presented herein would not have been possible.

\bibliographystyle{mnras}
\bibliography{mosdef_lris-bpt}

\begin{thebibliography}{}
\makeatletter
\relax
\def\mn@urlcharsother{\let\do\@makeother \do\$\do\&\do\#\do\^\do\_\do\%\do\~}
\def\mn@doi{\begingroup\mn@urlcharsother \@ifnextchar [ {\mn@doi@}
  {\mn@doi@[]}}
\def\mn@doi@[#1]#2{\def\@tempa{#1}\ifx\@tempa\@empty \href
  {http://dx.doi.org/#2} {doi:#2}\else \href {http://dx.doi.org/#2} {#1}\fi
  \endgroup}
\def\mn@eprint#1#2{\mn@eprint@#1:#2::\@nil}
\def\mn@eprint@arXiv#1{\href {http://arxiv.org/abs/#1} {{\tt arXiv:#1}}}
\def\mn@eprint@dblp#1{\href {http://dblp.uni-trier.de/rec/bibtex/#1.xml}
  {dblp:#1}}
\def\mn@eprint@#1:#2:#3:#4\@nil{\def\@tempa {#1}\def\@tempb {#2}\def\@tempc
  {#3}\ifx \@tempc \@empty \let \@tempc \@tempb \let \@tempb \@tempa \fi \ifx
  \@tempb \@empty \def\@tempb {arXiv}\fi \@ifundefined
  {mn@eprint@\@tempb}{\@tempb:\@tempc}{\expandafter \expandafter \csname
  mn@eprint@\@tempb\endcsname \expandafter{\@tempc}}}

\bibitem[\protect\citeauthoryear{{Abazajian} et~al.,}{{Abazajian}
  et~al.}{2009}]{Abazajian2009}
{Abazajian} K.~N.,  et~al., 2009, \mn@doi [\apjs]
  {10.1088/0067-0049/182/2/543}, \href
  {https://ui.adsabs.harvard.edu/abs/2009ApJS..182..543A} {182, 543}

\bibitem[\protect\citeauthoryear{{Asplund}, {Grevesse}, {Sauval}  \&
  {Scott}}{{Asplund} et~al.}{2009}]{Asplund2009}
{Asplund} M.,  {Grevesse} N.,  {Sauval} A.~J.,   {Scott} P.,  2009, \mn@doi
  [\araa] {10.1146/annurev.astro.46.060407.145222}, \href
  {https://ui.adsabs.harvard.edu/abs/2009ARA&A..47..481A} {47, 481}

\bibitem[\protect\citeauthoryear{{Baldwin}, {Phillips}  \&
  {Terlevich}}{{Baldwin} et~al.}{1981}]{Baldwin1981}
{Baldwin} J.~A.,  {Phillips} M.~M.,   {Terlevich} R.,  1981, \mn@doi [\pasp]
  {10.1086/130766}, \href
  {https://ui.adsabs.harvard.edu/abs/1981PASP...93....5B} {93, 5}

\bibitem[\protect\citeauthoryear{{Berg}, {Erb}, {Henry}, {Skillman}  \&
  {McQuinn}}{{Berg} et~al.}{2019}]{Berg2019}
{Berg} D.~A.,  {Erb} D.~K.,  {Henry} R. B.~C.,  {Skillman} E.~D.,   {McQuinn}
  K. B.~W.,  2019, \mn@doi [\apj] {10.3847/1538-4357/ab020a}, \href
  {https://ui.adsabs.harvard.edu/abs/2019ApJ...874...93B} {874, 93}

\bibitem[\protect\citeauthoryear{{Brinchmann}, {Pettini}  \&
  {Charlot}}{{Brinchmann} et~al.}{2008}]{Brinchmann2008}
{Brinchmann} J.,  {Pettini} M.,   {Charlot} S.,  2008, \mn@doi [\mnras]
  {10.1111/j.1365-2966.2008.12914.x}, \href
  {https://ui.adsabs.harvard.edu/abs/2008MNRAS.385..769B} {385, 769}

\bibitem[\protect\citeauthoryear{{Calzetti}, {Armus}, {Bohlin}, {Kinney},
  {Koornneef}  \& {Storchi-Bergmann}}{{Calzetti} et~al.}{2000}]{Calzetti2000}
{Calzetti} D.,  {Armus} L.,  {Bohlin} R.~C.,  {Kinney} A.~L.,  {Koornneef} J.,
   {Storchi-Bergmann} T.,  2000, \mn@doi [\apj] {10.1086/308692}, \href
  {https://ui.adsabs.harvard.edu/abs/2000ApJ...533..682C} {533, 682}

\bibitem[\protect\citeauthoryear{{Chabrier}}{{Chabrier}}{2003}]{Chabrier2003}
{Chabrier} G.,  2003, \mn@doi [\pasp] {10.1086/376392}, \href
  {https://ui.adsabs.harvard.edu/abs/2003PASP..115..763C} {115, 763}

\bibitem[\protect\citeauthoryear{{Chisholm}, {Rigby}, {Bayliss}, {Berg},
  {Dahle}, {Gladders}  \& {Sharon}}{{Chisholm} et~al.}{2019}]{Chisholm2019}
{Chisholm} J.,  {Rigby} J.~R.,  {Bayliss} M.,  {Berg} D.~A.,  {Dahle} H.,
  {Gladders} M.,   {Sharon} K.,  2019, \mn@doi [\apj]
  {10.3847/1538-4357/ab3104}, \href
  {https://ui.adsabs.harvard.edu/abs/2019ApJ...882..182C} {882, 182}

\bibitem[\protect\citeauthoryear{{Conroy}, {Gunn}  \& {White}}{{Conroy}
  et~al.}{2009}]{Conroy2009}
{Conroy} C.,  {Gunn} J.~E.,   {White} M.,  2009, \mn@doi [\apj]
  {10.1088/0004-637X/699/1/486}, \href
  {https://ui.adsabs.harvard.edu/abs/2009ApJ...699..486C} {699, 486}

\bibitem[\protect\citeauthoryear{{Crowther}, {Prinja}, {Pettini}  \&
  {Steidel}}{{Crowther} et~al.}{2006}]{Crowther2006}
{Crowther} P.~A.,  {Prinja} R.~K.,  {Pettini} M.,   {Steidel} C.~C.,  2006,
  \mn@doi [\mnras] {10.1111/j.1365-2966.2006.10164.x}, \href
  {https://ui.adsabs.harvard.edu/abs/2006MNRAS.368..895C} {368, 895}

\bibitem[\protect\citeauthoryear{{Cullen} et~al.,}{{Cullen}
  et~al.}{2019}]{Cullen2019}
{Cullen} F.,  et~al., 2019, \mn@doi [\mnras] {10.1093/mnras/stz1402}, \href
  {https://ui.adsabs.harvard.edu/abs/2019MNRAS.487.2038C} {487, 2038}

\bibitem[\protect\citeauthoryear{{Eldridge} \& {Stanway}}{{Eldridge} \&
  {Stanway}}{2012}]{Eldridge2012}
{Eldridge} J.~J.,  {Stanway} E.~R.,  2012, \mn@doi [\mnras]
  {10.1111/j.1365-2966.2011.19713.x}, \href
  {https://ui.adsabs.harvard.edu/abs/2012MNRAS.419..479E} {419, 479}

\bibitem[\protect\citeauthoryear{{Eldridge}, {Stanway}, {Xiao}, {McClelland },
  {Taylor}, {Ng}, {Greis}  \& {Bray}}{{Eldridge} et~al.}{2017}]{Eldridge2017}
{Eldridge} J.~J.,  {Stanway} E.~R.,  {Xiao} L.,  {McClelland } L.~A.~S.,
  {Taylor} G.,  {Ng} M.,  {Greis} S.~M.~L.,   {Bray} J.~C.,  2017, \mn@doi
  [\pasa] {10.1017/pasa.2017.51}, \href
  {https://ui.adsabs.harvard.edu/abs/2017PASA...34...58E} {34, e058}

\bibitem[\protect\citeauthoryear{{Erb}, {Shapley}, {Pettini}, {Steidel},
  {Reddy}  \& {Adelberger}}{{Erb} et~al.}{2006}]{Erb2006}
{Erb} D.~K.,  {Shapley} A.~E.,  {Pettini} M.,  {Steidel} C.~C.,  {Reddy} N.~A.,
    {Adelberger} K.~L.,  2006, \mn@doi [\apj] {10.1086/503623}, \href
  {https://ui.adsabs.harvard.edu/abs/2006ApJ...644..813E} {644, 813}

\bibitem[\protect\citeauthoryear{{Ferland} et~al.,}{{Ferland}
  et~al.}{2017}]{Ferland2017}
{Ferland} G.~J.,  et~al., 2017, Revista Mexicana de Astronomia y Astrofisica,
  \href {https://ui.adsabs.harvard.edu/abs/2017RMxAA..53..385F} {53, 385}

\bibitem[\protect\citeauthoryear{{Grogin} et~al.,}{{Grogin}
  et~al.}{2011}]{Grogin2011}
{Grogin} N.~A.,  et~al., 2011, \mn@doi [\apjs] {10.1088/0067-0049/197/2/35},
  \href {https://ui.adsabs.harvard.edu/abs/2011ApJS..197...35G} {197, 35}

\bibitem[\protect\citeauthoryear{{Halliday} et~al.,}{{Halliday}
  et~al.}{2008}]{Halliday2008}
{Halliday} C.,  et~al., 2008, \mn@doi [\aap] {10.1051/0004-6361:20078673},
  \href {https://ui.adsabs.harvard.edu/abs/2008A&A...479..417H} {479, 417}

\bibitem[\protect\citeauthoryear{{Kauffmann} et~al.,}{{Kauffmann}
  et~al.}{2003}]{Kauffmann2003}
{Kauffmann} G.,  et~al., 2003, \mn@doi [\mnras]
  {10.1111/j.1365-2966.2003.07154.x}, \href
  {https://ui.adsabs.harvard.edu/abs/2003MNRAS.346.1055K} {346, 1055}

\bibitem[\protect\citeauthoryear{{Kewley}, {Dopita}, {Sutherland}, {Heisler}
  \& {Trevena}}{{Kewley} et~al.}{2001}]{Kewley2001}
{Kewley} L.~J.,  {Dopita} M.~A.,  {Sutherland} R.~S.,  {Heisler} C.~A.,
  {Trevena} J.,  2001, \mn@doi [\apj] {10.1086/321545}, \href
  {https://ui.adsabs.harvard.edu/abs/2001ApJ...556..121K} {556, 121}

\bibitem[\protect\citeauthoryear{{Kewley}, {Dopita}, {Leitherer}, {Dav{\'e}},
  {Yuan}, {Allen}, {Groves}  \& {Sutherland}}{{Kewley}
  et~al.}{2013}]{Kewley2013}
{Kewley} L.~J.,  {Dopita} M.~A.,  {Leitherer} C.,  {Dav{\'e}} R.,  {Yuan} T.,
  {Allen} M.,  {Groves} B.,   {Sutherland} R.,  2013, \mn@doi [\apj]
  {10.1088/0004-637X/774/2/100}, \href
  {https://ui.adsabs.harvard.edu/abs/2013ApJ...774..100K} {774, 100}

\bibitem[\protect\citeauthoryear{{Kobayashi}, {Umeda}, {Nomoto}, {Tominaga}  \&
  {Ohkubo}}{{Kobayashi} et~al.}{2006}]{Kobayashi2006}
{Kobayashi} C.,  {Umeda} H.,  {Nomoto} K.,  {Tominaga} N.,   {Ohkubo} T.,
  2006, \mn@doi [\apj] {10.1086/508914}, \href
  {https://ui.adsabs.harvard.edu/abs/2006ApJ...653.1145K} {653, 1145}

\bibitem[\protect\citeauthoryear{{Kriek}, {van Dokkum}, {Labb{\'e}}, {Franx},
  {Illingworth}, {Marchesini}  \& {Quadri}}{{Kriek} et~al.}{2009}]{Kriek2009}
{Kriek} M.,  {van Dokkum} P.~G.,  {Labb{\'e}} I.,  {Franx} M.,  {Illingworth}
  G.~D.,  {Marchesini} D.,   {Quadri} R.~F.,  2009, \mn@doi [\apj]
  {10.1088/0004-637X/700/1/221}, \href
  {https://ui.adsabs.harvard.edu/abs/2009ApJ...700..221K} {700, 221}

\bibitem[\protect\citeauthoryear{{Kriek} et~al.,}{{Kriek}
  et~al.}{2015}]{Kriek2015}
{Kriek} M.,  et~al., 2015, \mn@doi [\apjs] {10.1088/0067-0049/218/2/15}, \href
  {https://ui.adsabs.harvard.edu/abs/2015ApJS..218...15K} {218, 15}

\bibitem[\protect\citeauthoryear{{Kriek} et~al.,}{{Kriek}
  et~al.}{2016}]{Kriek2016}
{Kriek} M.,  et~al., 2016, \mn@doi [\nat] {10.1038/nature20570}, \href
  {https://ui.adsabs.harvard.edu/abs/2016Natur.540..248K} {540, 248}

\bibitem[\protect\citeauthoryear{{Leitherer}, {Le{\~a}o}, {Heckman}, {Lennon},
  {Pettini}  \& {Robert}}{{Leitherer} et~al.}{2001}]{Leitherer2001}
{Leitherer} C.,  {Le{\~a}o} J. R.~S.,  {Heckman} T.~M.,  {Lennon} D.~J.,
  {Pettini} M.,   {Robert} C.,  2001, \mn@doi [\apj] {10.1086/319814}, \href
  {https://ui.adsabs.harvard.edu/abs/2001ApJ...550..724L} {550, 724}

\bibitem[\protect\citeauthoryear{{Liu}, {Shapley}, {Coil}, {Brinchmann}  \&
  {Ma}}{{Liu} et~al.}{2008}]{Liu2008}
{Liu} X.,  {Shapley} A.~E.,  {Coil} A.~L.,  {Brinchmann} J.,   {Ma} C.-P.,
  2008, \mn@doi [\apj] {10.1086/529030}, \href
  {https://ui.adsabs.harvard.edu/abs/2008ApJ...678..758L} {678, 758}

\bibitem[\protect\citeauthoryear{{Masters} et~al.,}{{Masters}
  et~al.}{2014}]{Masters2014}
{Masters} D.,  et~al., 2014, \mn@doi [\apj] {10.1088/0004-637X/785/2/153},
  \href {https://ui.adsabs.harvard.edu/abs/2014ApJ...785..153M} {785, 153}

\bibitem[\protect\citeauthoryear{{Masters}, {Faisst}  \& {Capak}}{{Masters}
  et~al.}{2016}]{Masters2016}
{Masters} D.,  {Faisst} A.,   {Capak} P.,  2016, \mn@doi [\apj]
  {10.3847/0004-637X/828/1/18}, \href
  {https://ui.adsabs.harvard.edu/abs/2016ApJ...828...18M} {828, 18}

\bibitem[\protect\citeauthoryear{{Matteucci}, {Spitoni}, {Romano}  \& {Rojas
  Arriagada}}{{Matteucci} et~al.}{2016}]{Matteucci2016}
{Matteucci} F.,  {Spitoni} E.,  {Romano} D.,   {Rojas Arriagada} A.,  2016, in
  Frontier Research in Astrophysics II (FRAPWS2016). p.~27

\bibitem[\protect\citeauthoryear{{McLean} et~al.,}{{McLean}
  et~al.}{2012}]{McLean2012}
{McLean} I.~S.,  et~al., 2012, in Ground-based and Airborne Instrumentation for
  Astronomy IV. p. 84460J, \mn@doi{10.1117/12.924794}

\bibitem[\protect\citeauthoryear{{Momcheva} et~al.,}{{Momcheva}
  et~al.}{2016}]{Momcheva2016}
{Momcheva} I.~G.,  et~al., 2016, \mn@doi [\apjs] {10.3847/0067-0049/225/2/27},
  \href {https://ui.adsabs.harvard.edu/abs/2016ApJS..225...27M} {225, 27}

\bibitem[\protect\citeauthoryear{{Nomoto}, {Tominaga}, {Umeda}, {Kobayashi}  \&
  {Maeda}}{{Nomoto} et~al.}{2006}]{Nomoto2006}
{Nomoto} K.,  {Tominaga} N.,  {Umeda} H.,  {Kobayashi} C.,   {Maeda} K.,  2006,
  \mn@doi [Nuclear Physics A] {10.1016/j.nuclphysa.2006.05.008}, \href
  {https://ui.adsabs.harvard.edu/abs/2006NuPhA.777..424N} {777, 424}

\bibitem[\protect\citeauthoryear{{Oke} et~al.,}{{Oke} et~al.}{1995}]{Oke1995}
{Oke} J.~B.,  et~al., 1995, \mn@doi [\pasp] {10.1086/133562}, \href
  {https://ui.adsabs.harvard.edu/abs/1995PASP..107..375O} {107, 375}

\bibitem[\protect\citeauthoryear{{Pilyugin}, {V{\'\i}lchez}, {Mattsson}  \&
  {Thuan}}{{Pilyugin} et~al.}{2012}]{Pilyugin2012}
{Pilyugin} L.~S.,  {V{\'\i}lchez} J.~M.,  {Mattsson} L.,   {Thuan} T.~X.,
  2012, \mn@doi [\mnras] {10.1111/j.1365-2966.2012.20420.x}, \href
  {https://ui.adsabs.harvard.edu/abs/2012MNRAS.421.1624P} {421, 1624}

\bibitem[\protect\citeauthoryear{{Rix}, {Pettini}, {Leitherer}, {Bresolin},
  {Kudritzki}  \& {Steidel}}{{Rix} et~al.}{2004}]{Rix2004}
{Rix} S.~A.,  {Pettini} M.,  {Leitherer} C.,  {Bresolin} F.,  {Kudritzki}
  R.-P.,   {Steidel} C.~C.,  2004, \mn@doi [\apj] {10.1086/424031}, \href
  {https://ui.adsabs.harvard.edu/abs/2004ApJ...615...98R} {615, 98}

\bibitem[\protect\citeauthoryear{{Sanders} et~al.,}{{Sanders}
  et~al.}{2016a}]{Sanders2016}
{Sanders} R.~L.,  et~al., 2016a, \mn@doi [\apj] {10.3847/0004-637X/816/1/23},
  \href {https://ui.adsabs.harvard.edu/abs/2016ApJ...816...23S} {816, 23}

\bibitem[\protect\citeauthoryear{{Sanders} et~al.,}{{Sanders}
  et~al.}{2016b}]{Sanders2016a}
{Sanders} R.~L.,  et~al., 2016b, \mn@doi [\apj] {10.3847/0004-637X/816/1/23},
  \href {https://ui.adsabs.harvard.edu/abs/2016ApJ...816...23S} {816, 23}

\bibitem[\protect\citeauthoryear{{Sanders} et~al.,}{{Sanders}
  et~al.}{2018}]{Sanders2018}
{Sanders} R.~L.,  et~al., 2018, \mn@doi [\apj] {10.3847/1538-4357/aabcbd},
  \href {https://ui.adsabs.harvard.edu/abs/2018ApJ...858...99S} {858, 99}

\bibitem[\protect\citeauthoryear{{Sanders} et~al.,}{{Sanders}
  et~al.}{2019}]{Sanders2019}
{Sanders} R.~L.,  et~al., 2019, \mn@doi [\mnras] {10.1093/mnras/stz3032}, \href
  {https://ui.adsabs.harvard.edu/abs/2019MNRAS.tmp.2653S} {p.~2653}

\bibitem[\protect\citeauthoryear{{Shapley}, {Steidel}, {Pettini}  \&
  {Adelberger}}{{Shapley} et~al.}{2003}]{Shapley2003}
{Shapley} A.~E.,  {Steidel} C.~C.,  {Pettini} M.,   {Adelberger} K.~L.,  2003,
  \mn@doi [\apj] {10.1086/373922}, \href
  {https://ui.adsabs.harvard.edu/abs/2003ApJ...588...65S} {588, 65}

\bibitem[\protect\citeauthoryear{{Shapley}, {Coil}, {Ma}  \& {Bundy}}{{Shapley}
  et~al.}{2005}]{Shapley2005}
{Shapley} A.~E.,  {Coil} A.~L.,  {Ma} C.-P.,   {Bundy} K.,  2005, \mn@doi
  [\apj] {10.1086/497630}, \href
  {https://ui.adsabs.harvard.edu/abs/2005ApJ...635.1006S} {635, 1006}

\bibitem[\protect\citeauthoryear{{Shapley}, {Steidel}, {Pettini}, {Adelberger}
  \& {Erb}}{{Shapley} et~al.}{2006}]{Shapley2006}
{Shapley} A.~E.,  {Steidel} C.~C.,  {Pettini} M.,  {Adelberger} K.~L.,   {Erb}
  D.~K.,  2006, \mn@doi [\apj] {10.1086/507511}, \href
  {https://ui.adsabs.harvard.edu/abs/2006ApJ...651..688S} {651, 688}

\bibitem[\protect\citeauthoryear{{Shapley} et~al.,}{{Shapley}
  et~al.}{2015}]{Shapley2015}
{Shapley} A.~E.,  et~al., 2015, \mn@doi [\apj] {10.1088/0004-637X/801/2/88},
  \href {https://ui.adsabs.harvard.edu/abs/2015ApJ...801...88S} {801, 88}

\bibitem[\protect\citeauthoryear{{Shapley} et~al.,}{{Shapley}
  et~al.}{2019}]{Shapley2019}
{Shapley} A.~E.,  et~al., 2019, \mn@doi [\apjl] {10.3847/2041-8213/ab385a},
  \href {https://ui.adsabs.harvard.edu/abs/2019ApJ...881L..35S} {881, L35}

\bibitem[\protect\citeauthoryear{{Shivaei} et~al.,}{{Shivaei}
  et~al.}{2016}]{Shivaei2016}
{Shivaei} I.,  et~al., 2016, \mn@doi [\apjl] {10.3847/2041-8205/820/2/L23},
  \href {https://ui.adsabs.harvard.edu/abs/2016ApJ...820L..23S} {820, L23}

\bibitem[\protect\citeauthoryear{{Sommariva}, {Mannucci}, {Cresci}, {Maiolino},
  {Marconi}, {Nagao}, {Baroni}  \& {Grazian}}{{Sommariva}
  et~al.}{2012}]{Sommariva2012}
{Sommariva} V.,  {Mannucci} F.,  {Cresci} G.,  {Maiolino} R.,  {Marconi} A.,
  {Nagao} T.,  {Baroni} A.,   {Grazian} A.,  2012, \mn@doi [\aap]
  {10.1051/0004-6361/201118134}, \href
  {https://ui.adsabs.harvard.edu/abs/2012A&A...539A.136S} {539, A136}

\bibitem[\protect\citeauthoryear{{Stanway} \& {Eldridge}}{{Stanway} \&
  {Eldridge}}{2018}]{Stanway2018}
{Stanway} E.~R.,  {Eldridge} J.~J.,  2018, \mn@doi [\mnras]
  {10.1093/mnras/sty1353}, \href
  {https://ui.adsabs.harvard.edu/abs/2018MNRAS.479...75S} {479, 75}

\bibitem[\protect\citeauthoryear{{Steidel} et~al.,}{{Steidel}
  et~al.}{2014}]{Steidel2014}
{Steidel} C.~C.,  et~al., 2014, \mn@doi [\apj] {10.1088/0004-637X/795/2/165},
  \href {https://ui.adsabs.harvard.edu/abs/2014ApJ...795..165S} {795, 165}

\bibitem[\protect\citeauthoryear{{Steidel}, {Strom}, {Pettini}, {Rudie},
  {Reddy}  \& {Trainor}}{{Steidel} et~al.}{2016}]{Steidel2016}
{Steidel} C.~C.,  {Strom} A.~L.,  {Pettini} M.,  {Rudie} G.~C.,  {Reddy} N.~A.,
    {Trainor} R.~F.,  2016, \mn@doi [\apj] {10.3847/0004-637X/826/2/159}, \href
  {https://ui.adsabs.harvard.edu/abs/2016ApJ...826..159S} {826, 159}

\bibitem[\protect\citeauthoryear{{Strom}, {Steidel}, {Rudie}, {Trainor},
  {Pettini}  \& {Reddy}}{{Strom} et~al.}{2017}]{Strom2017}
{Strom} A.~L.,  {Steidel} C.~C.,  {Rudie} G.~C.,  {Trainor} R.~F.,  {Pettini}
  M.,   {Reddy} N.~A.,  2017, \mn@doi [\apj] {10.3847/1538-4357/836/2/164},
  \href {https://ui.adsabs.harvard.edu/abs/2017ApJ...836..164S} {836, 164}

\bibitem[\protect\citeauthoryear{{Strom}, {Steidel}, {Rudie}, {Trainor}  \&
  {Pettini}}{{Strom} et~al.}{2018}]{Strom2018}
{Strom} A.~L.,  {Steidel} C.~C.,  {Rudie} G.~C.,  {Trainor} R.~F.,   {Pettini}
  M.,  2018, \mn@doi [\apj] {10.3847/1538-4357/aae1a5}, \href
  {https://ui.adsabs.harvard.edu/abs/2018ApJ...868..117S} {868, 117}

\bibitem[\protect\citeauthoryear{{Veilleux} \& {Osterbrock}}{{Veilleux} \&
  {Osterbrock}}{1987}]{Veilleux1987}
{Veilleux} S.,  {Osterbrock} D.~E.,  1987, \mn@doi [\apjs] {10.1086/191166},
  \href {https://ui.adsabs.harvard.edu/abs/1987ApJS...63..295V} {63, 295}

\bibitem[\protect\citeauthoryear{{Weinberg}, {Andrews}  \&
  {Freudenburg}}{{Weinberg} et~al.}{2017}]{Weinberg2017}
{Weinberg} D.~H.,  {Andrews} B.~H.,   {Freudenburg} J.,  2017, \mn@doi [\apj]
  {10.3847/1538-4357/837/2/183}, \href
  {https://ui.adsabs.harvard.edu/abs/2017ApJ...837..183W} {837, 183}

\bibitem[\protect\citeauthoryear{{Zahid}, {Kudritzki}, {Conroy}, {Andrews}  \&
  {Ho}}{{Zahid} et~al.}{2017}]{Zahid2017}
{Zahid} H.~J.,  {Kudritzki} R.-P.,  {Conroy} C.,  {Andrews} B.,   {Ho} I.~T.,
  2017, \mn@doi [\apj] {10.3847/1538-4357/aa88ae}, \href
  {https://ui.adsabs.harvard.edu/abs/2017ApJ...847...18Z} {847, 18}

\makeatother
\end{thebibliography}

\end{document}